\documentclass[aps,floatfix,prd,showpacs]{revtex4}
\usepackage{graphicx}% Include figure files
\usepackage{dcolumn}% Align table columns on decimal point
\usepackage{bm}% bold math
\usepackage{graphics}
\usepackage{amssymb}
\usepackage{rotating,multirow}
\usepackage{amsthm,amsmath}
\usepackage{array}
\usepackage[all]{xy}
\newtheorem{theorem}{Theorem}[section]
\newtheorem{lemma}[theorem]{Lemma}

\theoremstyle{definition}
\newtheorem{definition}[theorem]{Definition}

\theoremstyle{remark}

\begin{document}

\title{Affine Kac-Moody symmetric spaces associated with untwisted Kac-Moody algebras}
\author{Saudamini Nayak, S.S.Rout and K.C.Pati}
\email{kcpati@nitrkl.ac.in}
\affiliation{
Department of Mathematics,
National Institute of Technology, Rourkela,
Odisha-769 008 India.}
\begin{abstract}
In this paper we have computed all the affine Kac-Moody symmetric spaces which are tame Fr$\acute{e}$chet manifolds starting from the Vogan diagrams related to the affine untwisted Kac-Moody algebras. The detail computation of affine Kac-Moody symmetric spaces associated with $A_{1}^{(1)}$ and $A_{2}^{(1)}$ are shown algebraically to corroborate our method.
\end{abstract}
\maketitle
\vspace*{-1.5cm}

\section{Introduction}
Finite dimensional symmetric spaces \cite{Loos 1969, Helgason2001} are rather well understood mathematical objects which have recently gained much importance in both mathematics and physics due to their intimate connections with random matrix theories, Reimannian geometries and their applications to many integrable systems, quantum transport phenomena (disordered system etc) \cite{Caselle,Caselle2006,Dyson1970,Olshanetsky1983}. A compact irreducible symmetric space is either a compact simple Lie group $G$ or a quotient $G/K$ of a compact simple Lie group by the fixed point set of an involution $\rho$ (or an open subgroup of it) and $g=t \oplus p$ is the decomposition of the Lie algebra $g$ of group $G$ into +1 and -1 eigenvalue spaces of $\rho$ then $K$ acts on $g$ by adjoint representation leaving the decomposition invariant. The restriction of this action to $p$ can be identified with the isotropy representation of $G/K$ and we know that the isotropy representation of a symmetric space is polar.
\par
With the advent of Kac-Moody algebras \cite{Kac1968,Kac1990, Moody1967} which can be considered as the generalization of finite dimensional Lie algebras, naturally a search of infinite dimensional version of symmetric spaces began, the closest generalization being the affine Kac-Moody symmetric spaces. An affine Kac-Moody symmetric space is by definition either an affine Kac-Moody group $\hat{G}$(group type) or a quotient $\hat{G}/\hat{G}^{\hat{\rho}}$ of $\hat{G}$ by the fixed point set of an involution of the second kind. In fact if $\hat{g}=\hat{t}+ \hat{p}$ in the splitting of the Lie algebra of $\hat{G}$
into the $\pm1$ eigen spaces of $\hat{\rho}$. Then the metric of $\hat{G}/\hat{K}$ is the left invariant metric obtained from the restriction of the inner product of $\hat{g}$ to $\hat{p}$. In finite dimension, the isotropy representation is polar itself while in infinite dimension it leaves invariant a co-dimension-2 sub-manifold which can be identified with a (pre-)Hilbert space and the induced action on this space is a polar action by affine isometries.
\par
The study on affine Kac-Moody symmetric spaces began with C.L.Terng \cite{Terng1995} who conjectured the existence of infinite dimensional symmetric spaces. Important progresses towards their constructions and geometries are achieved by B. Popescu \cite{Popescu2005}, W. Freyn \cite{Freyn2009}, Heintze \cite{Heintze2006,Heintze2012}, et al. Now it has been shown that affine Kac-Moody symmetric spaces are tame Fr$\acute{e}$chet manifolds. In particular let $G$ be the simply connected Lie group with Lie algebra $g$ and denote $ \sigma $ be the automorphism of $G$ corresponding to $\sigma \in Aut(g)$. Then the loop group 
\begin{equation}
L(G, \sigma)=\{g: \mathbb{R}\rightarrow G\mid g \in \mathcal{C}^{\infty}, g(t+2\pi)=\sigma(g(t)) \;\; \forall t\}
\end{equation} with point wise multiplication is a Fr$\acute{e}$chet Lie group with Lie algebra $L(g, \sigma)$. The affine Kac-Moody group $\hat{L}(G, \sigma)$ will be a $T^{2}=S^{1}\times S^{1}$ bundle over $L(G, \sigma)$. By construction $\hat{L}(G, \sigma)$ is a Fr$\acute{e}$chet group and it is a Lorentz manifold. It is also well known that $\hat{L}(G, \sigma)/\hat{L}(G, \sigma)^{\hat{\rho}}$ for any involution $\hat{\rho}$ of $\hat{L}(G, \sigma)$ and $L(G, \sigma)/L(G, \sigma)^{\rho}$ for any involution $\hat{\rho}$ are tame Fr$\acute{e}$chet.
\par
The classification of affine Kac-Moody symmetric spaces is essentially equivalent to the classification of involutions of affine Kac-Moody algebras upto conjugation. This has been achieved by a long series of papers by various authors Batra \cite{Batra2000}, Levstein \cite{Levstein1988}, Kobayashi \cite{Kobayashi1986}, Rousseau and Messaoud  \cite{Rousseau2003}. There is a one-to-one correspondence between real forms of Kac-Moody algebras, involutions. The diagramatic representation of real forms are the Vogan diagrams (to each real form there is a unique Vogan diagram) which are Dynkin diagrams together with some extra pieces of informations. We have collected/construted the Vogan diagrams related with untwisted classical algebras $A_{n}^{(1)},B_{n}^{(1)}, C_{n}^{(1)}, D_{n}^{(1)}$ explicitly. From the Vogan diagrams we have obtained the fixed point set as well as the real forms of the algebras and then we have constructed affine Kac-Moody symmetric spaces.
\par
In chapter-2 we have given a brief introduction to affine untwisted Kac-Moody algebras with their involutions, real forms and Vogan diagrams.
\par
In chapter-3 we have explicitly calculated the affine Kac-Moody symmetric spaces with two elementary untwisted Kac-Moody algebras $A_{1}^{(1)}$ and $A_{2}^{(1)}$ relating them with their Vogan diagrams to corroborate our technique to construct affine Kac-Moody symmetric spaces. Towards the end of the chapter we have given exhaustive list of all the affine Kac-Moody symmetric spaces together with their real forms and fixed algebras. Chapter-4 contains appendix. 
\section{PRELIMINARIES}\label{PRELIM}

\subsection{Kac-Moody Lie algebras}
Let $I=[1, n+1], n \in \mathbb{N}$, be an interval in $\mathbb{N}$.
A matrix $A=(a_{i j})_{i, j \in I}$ with integer coefficients is
called a generalized Cartan matrix if it satisfies the following
conditions:
\begin{enumerate}
\item $a_{i i}=2$ for $i=1,\cdots,n+1$.
\item $a_{i j}\leq 0$  for $i\neq j$.
\item $a_{i j}=0$ iff $a_{j i}=0$.
\end{enumerate}
A realization of $A$ is a triple $(\mathfrak{h}, \pi, \check{\pi})$, where $\mathfrak{h}$ is a finite-dimensional complex vector space and $\pi= \{\alpha_{i}\}_{i\in I} \subset \mathfrak{h}^{\ast}$ and $\check{\pi}=\{\alpha_{i}\}_{i\in I} \subset \mathfrak{h}$ are indexed subsets in $\mathfrak{h}^{\ast}$ and $\mathfrak{h}$ respectively, and they satisfy
\begin{enumerate}
\item both $\pi, \check{\pi}$ are linearly independent.
\item $<\alpha_{i}, \check{\alpha_{j}}>=a_{j i}$, for $i,j= 1,\cdots,n$.
\item rank $(A)= 2n$ - dim $\mathfrak{h}$.
\end{enumerate}
For any $n \times n$ matrix $A$ there exits a unique(upto isomorphism) realization.
\par
 Given two matrices $A$ and $A^{\prime}$ and their realizations $(\mathfrak{h}, \pi, \check{\pi})$ and $(\mathfrak{h}^{\prime}, \pi^{\prime}, \check{\pi}^{\prime})$, we obtain a realization of the direct sum of two matrices $(\mathfrak{h} \oplus \mathfrak{h}^{\prime}, \pi \otimes \{0\}\cup \{0\} \otimes \pi^{\prime}, \check{\pi} \otimes \{0\}\cup \{0\} \otimes \check{\pi}^{\prime})$ which is called direct sum of the realizations.
 \par
 A matrix $A$ is called decomposable if after reordering of indices $A$ decomposes into a non-trivial direct sum. Otherwise $A$  is called indecomposable. 

%\begin{definition}[Definition]
Let $A=(a_{i j})$ be a generalized Cartan matrix and let $(\mathfrak{h},\pi, \check{\pi})$ be a realization of $A$. Let $g(A)$ be a complex Lie algebra with generators $e_{i}, f_{i}$ for $i=1 \cdots n$ and $\mathfrak{h}$ and the following defining relations:
\begin{eqnarray*}\label{a1}
{[e_{i}, f_{j}]} &=&\delta_{ij}\check{\alpha_{i}}\\
 {[h, h']} &=&0 \\
  {[h, e_{i}]} &=&<\alpha_{i}, h>e_{i}\\
   {[h, f_{i}]} &=&-<\alpha_{i}, h>f_{i}
   \end{eqnarray*}and
  the Serre relations
  \begin{equation}\label{eq1}
  (ad \ e_{i})^{1-{a}_{ij}}e_{j}=0,\quad (ad\ f_{i})^{1-{a}_{ij}}f_{j}=0;~~ \forall i\neq
  j,
  \end{equation}
  The Lie algebra $g=g(A)$ is called a Kac-Moody algebra. The subalgebra $\mathfrak{h}$ of $g$ is called the Cartan subalgebra. The matrix $A$ is the Cartan matrix of $g$ which is of rank $n$. 
The elements $e_{i}, f_{i}$ for $i=1 \cdots n$, are called Chevalley generators and they generates the subalgebra $g^{\prime}=[g, g]$ and $g= g^{\prime} + \mathfrak{h}$.
\subsection{Affine Kac-Moody Algebras}
Consider the generalized Cartan matrix $A$. It is called a Cartan matrix of affine type if
\begin{itemize}
  \item $A$ is an indecomposable matrix, i.e. after the indices are reordered $A$ cannot be written in the form
$\begin{pmatrix}
A_{1} & 0 \\
0 & A_{2}
\end{pmatrix}.$
\item There exits a vector $(a_{i})_{i=1}^{n+1}$, with $a_{i}$ all positive such that $A(a_{i})_{i=1}^{n+1}=0$.
\end{itemize}
Then, the algebra $g$ associated with $A$ is called an affine Kac-Moody algebra. Affine Kac-Moody algebras are of two types untwisted and twisted. In this paper we have confined ourselves to untwisted case only.
\subsection{A realization of  non-twisted affine Kac-Moody Lie algebra}
Let $L=\mathcal{C}[t, t^{-1}]$ be the algebra of Laurent polynomials in $t$. The residue of the Laurent polynomial $P= \sum _{j \in \mathbb{Z}}c_{j}t^{j}$(where all but finite number of $c_{j}$ are zero) is  Res$P=c_{-1}$. 
\par
Let $\mathring{g}$ be a finite-dimensional simple Lie algebra over $\mathbb{C}$ of type $X_{n}$, then $L(\mathring{g})= L\otimes \mathring{g}$ is an infinite-dimensional Lie algebra with the bracket 
\begin{equation}
[P \otimes X, Q \otimes Y]= P Q \otimes [x, y]~~~~
P,Q \in L; X,Y \in \mathring{g}.
\end{equation}
Fix a non-degenerate, invariant, symmetric bilinear form $(.,.)$ in $\mathring{g}$ and extend this form to an $L$ valued form  $(.,.)_{t}$ on $L(\mathring{g})$ by
\begin{equation}
(P \otimes x, Q \otimes y)_{t}= P Q(x, y) ~~~~ P, Q \in L; x,y \in \mathring{g}. 
\end{equation}
The derivation $t^{j}(d/dt)$ of $L$ extends to $L(\mathring{g})$ by
\begin{equation}
t^{j} \frac{d}{d t}(P \otimes X)= t^{j}\frac{d P}{d t} \otimes X, ~~~~P \in L; X \in \mathring{g}.
\end{equation}
Therefore $\psi(a, b)= Res(\frac{d a}{d t}, b)_{t}$ for $a, b \in L(\mathring{g})$ defines a two-cocycle on $L(\mathring{g})$.
\par
Now we denote by $\tilde{L}(\mathring{g})$ the central extension of the Lie algebra $L(\mathring{g})$ associated to the cocycle $\psi$. Explicitly $\tilde{L}(\mathring{g})= L(\mathring{g})\oplus \mathbb{C}c$ with the bracket
\begin{equation}
[a+ \lambda c, b+ \mu c]= [a, b]+ \psi(a,b)c \;\;a, b \in L(\mathring{g}); \lambda, \mu \in \mathbb{C}.
\end{equation}
Finally, denote by $\hat{L}(\mathring{g})$ is the Lie algebra which is obtained by adjoining to $\tilde{L}(\mathring{g})$ a derivation $d$ which acts on $L(\mathring{g})$ as $ t \frac{d}{d t}$ and kills $c$. Explicitly we have $\hat{L}(\mathring{g})= L(\mathring{g})\oplus \mathbb{C}c \oplus \mathbb{C}d$ with the bracket defined by 
\begin{equation}
[t^{k}\otimes x + \lambda c+ \mu d, t^{j}\otimes y+ \lambda_{1}c+ \mu d]= t^{j+k} \otimes [x, y]+ \mu j t^{j} \otimes y- \mu_{1} k t^{k} \otimes x+ k \delta_{j, -k}(x, y)c,
\end{equation}
where $x,y \in \mathring{g}; \lambda, \mu, \lambda_{1}, \mu_{1} \in \mathbb{C}; j, k \in \mathbb{Z}$. This $\hat{L}(\mathring{g})$ is a non- twisted affine Kac-Moody Lie algebra associated to the affine matrix $A$ of type $X_{n}^{(1)}$. 
\subsection{Automorphisms and Real forms of non-twisted affine Kac-Moody algebras}
Define a group $G$ acting on the algebra $g$ through adjoint representation $Ad: G\rightarrow Aut{(g)}$. It is generated by the subgroup $U_{\alpha}$ for $\alpha \in \pm \pi$ and $Ad(U_{\alpha})= exp({{ad}(g_{\alpha})})$. 
\par
A maximal $ad_{g}$-diagonalizable subalgebra of $g$ is called a Cartan subalgebra. Every Cartan subalgebra of $g$ is $Ad(G)$-conjugate to the standard Cartan subalgebra $\mathfrak{h}$. A Borel subalgebra of $g$ is maximal completely solvable subalgebra. It is conjugated by $Ad(G)$ to $b^{+}$ and $b^{-}$ where $b^{+}=\mathfrak{h}\oplus \bigoplus_{\alpha>0}g_{\alpha}$ and $b^{-}=\mathfrak{h}\oplus\bigoplus_{\alpha<0}g_{\alpha}$. However $b^{+}$ and $b^{-}$ are not conjugated under $Ad(G)$. So there are two conjugacy classes of Borel subalgebra the positive and negative subalgebras. 
\par
A real form of $g$ is a algebra $g_{\mathbb{R}}$ over $\mathbb{R}$ such that there exists an isomorphism from $g$ to $g_{\mathbb{R}}\otimes \mathbb{C}$. If we replace $\mathbb{C}$ by $\mathbb{R}$ in definition of $g$ then we obtain a real form of $g_{\mathbb{R}}$ which is called split real form. 
\par
An automorphism $\sigma$ of $g$ is called an involution if $\sigma^{2}=Id$. The involution is called semi-linear if $\sigma(\lambda x)= \bar{\lambda} \sigma(x)$ for $\lambda \in \mathbb{C}$ and $x \in g$. A real form of $g$ correspondences to a semi-linear involution of $g$. A linear or semi-linear automorphism $\sigma$ of $g$ is said to be of first kind if $\sigma(b^{+})$ is $Ad(G)$- conjugate to $b^{+}$ and it is of second kind if $\sigma(b^{+})$ is $Ad(G)$-conjugate to $b_{-}$. Any automorphism of $g$ is either an automorphism of first kind (type 1) or an automorphism of second kind (type 2). 
\par
Let $g_{\mathbb{R}}$ be a real form of $g$. Fix an isomorphism from $g$ to $g_{\mathbb{R}}\otimes \mathbb{C}$. Then the Galois group $\Gamma= Gal(\mathbb{C}/\mathbb{R})$ acts on $g$  and the corresponding group $G$. Then $g_{\mathbb{R}}$ can be identified with the fixed point set $g^{\Gamma}$. If $\Gamma$ consists of first kind automorphism then we say $g_{\mathbb{R}}$ is almost split, otherwise if the non-trivial element of $\Gamma$ is of second kind automorphism then we say $g_{\mathbb{R}}$ is almost compact(non-compact). Denote the group of $\mathbb{C}$-linear or semilinear automorphisms of $g$ as $Aut_{\mathbb{R}}(g)$. The group $Aut(g)$ is normal in $Aut_{\mathbb{R}}(g)$ and of index 2. A semilinear automorphism of order 2 of $g$ is called a semiinvolution of $g$.
\begin{definition}
Let $\sigma^{\prime}$ be a semi-involution of $g$ of second kind and let $g_{\mathbb{R}}= g^{\sigma^{\prime}}$ be the corresponding almost compact real form. A Cartan semi-involution $\omega^{\prime}$ which commutes with $\sigma^{\prime}$ is called a Cartan semi-involution for $\sigma^{\prime}$ or $g_{\mathbb{R}}$. The involution $\sigma=\sigma^{\prime}\omega^{\prime}$ is called a Cartan involution of $\sigma^{\prime}$ and also the restriction $\omega_{\mathbb{R}^{\prime}}$ of $\sigma$ to $g_{\mathbb{R}}$ is called Cartan involution for $g_{\mathbb{R}}$.  
\end{definition}
The algebra of fixed points $\mathfrak{t_{0}}= g_{\mathbb{R}}^{\sigma}$ is called a maximal compact subalgebra of $g_{\mathbb{R}}$. Now we have the Cartan decomposition $g_{\mathbb{R}}= \mathfrak{t_{0}} \oplus \mathfrak{p_{0}}$ and $\mathfrak{t_{1}}=\mathfrak{t_{0}} \oplus i\mathfrak{p_{0}}$ where $\mathfrak{p_{0}}$ is the eigenspace of $\omega_{\mathbb{R}}^{\prime}$ for eigen value $-1$. Let $t_{0}$ be a maximal abelian subspace of $\mathfrak{t_{0}}$. Then $\mathfrak{h_{0}}=Z_{g_{\mathbb{R}}}(t_{0})$ is a $\sigma$-stable Cartan subalgebra of the almost compact real form $g_{\mathbb{R}}$ of the form $\mathfrak{h_{0}}=t_{0}\oplus a_{0}$ with $a_{0}\subseteq \mathfrak{p_{0}}$.
\begin{definition}
A $\sigma$-stable Cartan subalgebra $ \mathfrak{h_{0}}=t_{0} \oplus a_{0}$ with $t_{0}\subseteq \mathfrak{t_{0}}$ and $a_{0} \subseteq \mathfrak{p_{0}}$ of an almost compact real form $g_{\mathbb{R}}$ is maximally compact if the dimension of $t_{0}$ is as large as possible and it is maximally non-compact if the dimension of $a_{0}$ is as large as possible.
\end{definition}
A maximally compact Cartan subalgebra $\mathfrak{h_{0}}$ of an almost compact real form $g_{\mathbb{R}}$ has the property that all the roots are real on $a_{0}$ and imaginary on $t_{0}$. One says that a root is real if it takes real value on $\mathfrak{h}_{0}= t_{0} \oplus a_{0}$, i.e. vanishes on $t_{0}$. It is imaginary if it takes imaginary value on $\mathfrak{h_{0}}$, i.e. vanishes on $a_{0}$ and complex otherwise.
\subsection{Classification of Real forms}
Under $Aut{(g)}$ there is a one-one correspondence between the conjugacy classes of involutions(linear) of second kind of $g$ and the conjugacy classes of almost split real forms of $g$. Again there is a bijection between the conjugacy classes under $Aut{(g)}$ of semi-involution of second kind and conjugacy classes of involution of first kind. Thus one obtain under $Aut{(g})$ a one-to-one correspondence between conjugacy classes of linear involutions of first kind(including identity) and the the conjugacy classes of almost compact real forms of $g$. The compact real form is unique and corresponds to  the identity.

\par{} 
Let $\mathfrak{h_{0}}$ be a $\sigma$-stable Cartan subalgebra of $g_{\mathbb{R}}$.Then there are no real roots iff $\mathfrak{h_{0}}$ is maximally compact.
\par{}
Let $g_\mathbb{R}$ be almost compact real form of $g$ corresponding to the semi-involution of the second kind $\sigma^{\prime}$ of $g$. Let $\sigma$ be the Cartan involution of $g_{\mathbb{R}}$ and let $g_{\mathbb{R}}=\mathfrak{t_{0}}\oplus \mathfrak{p_{0}}$ be the corresponding Cartan decomposition\cite{Birkhoff1937,Gilmore,Jacobson1979}. Let $t_{0}$ be maximal abelian subspace of $\mathfrak{t_{0}}$. Then $\mathfrak{h_{0}}=Z_{g_{\mathbb{R}}}(t_{0})$ is a $\sigma$-stable Cartan subalgebra of $g_{\mathbb{R}}$ of the form $\mathfrak{h}_{0}=t_{0}\oplus a_{0}$ with $a_{0}\subseteq \mathfrak{p_{0}}$. This $\mathfrak{h_{0}}$ is a maximally Cartan subalgebra of $g_{\mathbb{R}}$ because $t_{0}$ is as large as possible.
\par{}
For any root $\alpha, \;\; \sigma(\alpha) $ is the root $\sigma \alpha(H)= \alpha(\sigma^{-1}H)$. If $\alpha$  is imaginary then $\sigma(\alpha)=\alpha$ and $ \alpha $ vanishes on $a_{0}$. Thus $g_{\alpha}$ is $\sigma$-stable and we have $g_{\alpha}= (g_{\alpha}\cap \mathfrak{t})\oplus(g_{\alpha}\cap \mathfrak{p}) $. Again $\dim(g_{\alpha})=1$, so $g_{\alpha}\subseteq \mathfrak{t}$ or $g_{\alpha}\subseteq \mathfrak{p}$. An imaginary root $\alpha$ is called compact if $g_{\alpha}\subseteq \mathfrak{t}$ and is non-compact if $g_{\alpha}\subseteq \mathfrak{p}$.
\begin{theorem}(Theorem 45 \cite{Freyn2009})
Let $g$ be an complex affine Kac-Moody algebra and $\mathcal{C}$ be a real form of it which is compact type. The conjugacy classes of real forms of non compact type of $g$ are in bijection with the conjugacy classes of involutions on $\mathcal{C}$. The correspondence is given by $\mathcal{C}=K\oplus P\mapsto K\oplus iP$ where $K$ and $P$ are the $\pm$-eigen spaces for the involution.
\end{theorem}
However every real form is either of compact type or of non-compact type, a mixed type is not possible.
\begin{lemma}(Lemma 47 \cite{Freyn2009})
Let $g_{\mathbb{R}}$ be a real form of non-compact type. Let $g_{\mathbb{R}}= K \oplus P$ be Cartan decomposition. The Cartan Killing form is negative definite on $K$ and positive definite on $P$.
\end{lemma}
\subsection{Vogan diagrams}
For classification of real forms of affine Kac-Moody algebra there are two main approaches: One focuses the maximal non-compact Cartan subalgebra that leads to Satake diagrams \cite{Tripathy2006}. The other one is on maximal compact Cartan subalgebra that leads to Vogan diagrams \cite{Knapp2002, Batra2002, Chuah2004}.
\par
Let almost compact real form $g_{\mathbb{R}}$ of $g$ and $\sigma$ be the Cartan involution on $g_{\mathbb{R}}$ leading to the Cartan decomposition $g_{\mathbb{R}}=\mathfrak{t_{0}} \oplus \mathfrak{p_{0}}$. Let $\mathfrak{h_{0}}$ be the maximally compact $\sigma$-stable Cartan subalgebra of $g_{\mathbb{R}}$
with complexification $\mathfrak{h}= \mathfrak{t} \oplus \mathfrak{a}$. Let us denote $\Delta= \Delta(g, \mathfrak{h})$ be the set of roots of $g$ with respect to $\mathfrak{h}$. This set doesn't contain any real root as $\mathfrak{h_{0}}$ is assumed to be maximally compact. From $\Delta$ we choose a positive system $\Delta^{+}$ that takes $i t_{0}$ before $a$. since $\sigma$ is $+1$ on $\mathfrak{t_{0}}$ and $-1$ on $a_{0}$ and since there are no real roots $\sigma(\Delta^{+})=\Delta^{+}$. Therefore $\sigma$ permutes the simple roots. It fixes the simple roots that are imaginary and permutes in 2-cycles the simple roots that are complex.
\begin{definition}
By Vogan diagram of the triple $(g_{\mathbb{R}}, \mathfrak{h_{0}}, \Delta^{+})$ we mean the Dynkin diagram of $\Delta^{+}$ with the 2-element orbits under $\sigma$  labelled an arrow and with the 1-element orbit painted or not depending upon whether the corresponding imaginary simple root is non-compact or compact.
\end{definition}
Every Vogan diagram represents an almost compact(non-compact) real form of some affine Kac-Moody Lie algebra. Two diagrams may represent isomorphic algebras and in that case the diagrams are equivalent. So the classification of Vogan diagram gives rise to the classification of almost compact real form of affine Kac-Moody Lie algebra.
\par
The equivalence of Vogan diagram is defined as the equivalence relation generated by the following two operations:
\begin{enumerate}
\item Applications of an automorphism of the Dynkin diagram.
\item Change in the positive system by reflection in a simple, non-compact root, i.e. by a vertex which is colored in the Vogan diagram.
\end{enumerate}
As a consequence of reflection by a simple non-compact root $\alpha$, the rules for single and triple lines is that we have $\alpha$ colored and its immediate neighbour is changed to the opposite color. The rule for double line is that if $\alpha$ is the smaller root, then there is no change in the color of immediate neighbour, but we leave $\alpha$ colored. If $\alpha$ is a bigger root, then we leave $\alpha$ colored and the immediate neighbour is changed to the opposite color. 
\par
If two Vogan diagrams aren't equivalent to each other, then they are called non-equivalent.
\begin{definition}
An abstract Vogan diagram is an irreducible abstract Dynkin diagram of non-twisted affine Kac-Moody Lie algebra with two additional piece of structure as follows:
\begin{enumerate}
\item One is an automorphism of order 1 or 2 of the diagram, which is indicated by labelling the 2-element orbits.
\item Second one is a subset of 1-element orbits which is to be indicated by pointing the vertices corresponding to the members of the subset. 
\end{enumerate}
\end{definition}
Every Vogan diagram is an abstract Vogan diagram. 
It is always convenient to represent equivalence class of Vogan diagrams with minimum number of vertices painted. We have Borel Seibenthal theorem for affine Kac-Moody algebras \cite{Chuah2006} which states that
\begin{theorem}
Every equivalence class of Vogan diagram has a representative with atmost two vertices painted.
\end{theorem}
Some more important results: 
\begin{theorem}
If an abstract Vogan diagram for an non-twisted affine Kac-Moody Lie algebra is given, then there exits an almost compact real form of a non-twisted affine Kac-Moody Lie algebra such that the given diagram is the Vogan diagram of this almost compact real form.
\end{theorem}
\begin{theorem}
If two almost compact real forms of a non-twisted affine Kac-Moody Lie algebra $g$ have equivalent Vogan diagram then they are isomorphic.
\end{theorem}
\section{Affine Kac-Moody symmetric space}
In this chapter we briefly review with the definition and geometry of the affine Kac-Moody symmetric spaces \cite{Freyn2009,Heintze2006,Popescu2005} with explicit determination of affine Kac-Moody symmetric spaces associated with $A_{1}^{(1)}$ and $A_{2}^{(1)}$. 
\begin{definition}
A tame Fr$\acute{e}$chet manifold $M$ with a weak metric having a Levi-civita connection is called a symmettric space, iff $\forall p \in M$ there is an involution isometry $\rho_{p}$, such that $p$ is an isolated fixed point of $\rho_{p}$.
\end{definition}
\begin{definition}
An(affine) Kac-Moody symmetric space $M$ is a tame Fr$\acute{e}$chet Lorentz symmetric space such that its isometry group $I(M)$ contains a transitive subgroup isomorphic to an affine geometric Kac-Moody group $\mathcal{H}$ and the intersection of the isotropy group of a point with $\mathcal{H}$ is a loop group of compact type.
\end{definition}
\begin{theorem}(Affine Kac-Moody symmetric spaces of compact type)\newline
Both the Kac-Moody group $\widehat{MG}_{\mathbb{R}}^{\sigma}$ equipped with its Ad-invariant metric, and the quotient space $X=\widehat{MG}_{\mathbb{R}}^{\sigma}/{Fix(\rho_{*})}$  equipped with its $Ad(Fix(\rho_{*}))$-invariant metric are tame  Fr$\acute{e}$chet symmetric spaces of the compact type with respect to their Ad-invariant metric. Their curvatures satisfy $\big<R(X,Y)X,Y \big>\geq 0.$
\end{theorem}
\begin{theorem}(Affine Kac-Moody symmetric spaces of non-compact type)\newline
Both quotient spaces $X=\widehat{MG}_{\mathbb{C}}^{\sigma}/\widehat{MG}_{\mathbb{R}}^{\sigma}$   and $X=H/{Fix(\rho_{*})}$ where $H$ is a non-compact real form of $\widehat{MG}_{\mathbb{C}}^{\sigma}$ with their Ad-invariant metric are tame  Fr$\acute{e}$chet symmetric spaces of non-compact type. Their curvatures satisfy $\big<R(X,Y)X,Y \big>\leq 0.$ Furthermore Kac-Moody symmetric spaces of the non compact type are diffeomorphic to vector space. 
\end{theorem}
\begin{theorem}(Duality)\newline
Affine Kac-Moody symmetric spaces of compact type are dual to the Affine Kac-Moody symmetric spaces of non-compact type and vice versa.
\end{theorem}
\par
We can summarize all the results we have discussed so far as follows:\\
There is an one-one correspondence between the conjugacy classes of involution of second kind of affine Kac-Moody algebra $g$ and almost split real form and also  between the conjugacy classes of involution of first kind and almost compact real form of $g$. Again to each real form there is a unique Vogan diagram. On the other hand all real forms are of two types: compact, non-compact and the conjugacy classes of real forms of non-compact type of $g$ are in bijection with the conjugacy classes of involution on the compact real form ($\mathcal{C}$). So the affine  Kac-Moody symmetric spaces can be classified using compact real form and involution of second kind. Now we can conclude that classification affine  Kac-Moody symmetric spaces are intimately linked with classification of Vogan diagram.  Starting with the compact real form $\mathcal{C}$ we can construct all the various non-compact real forms by applying involutive automorphisms to $\mathcal{C}$ followed by Weyl unitary trick.
\subsection{Affine Kac-Moody symmetric spaces associated with $A_{1}^{(1)}$}
The chevelley generators for $A_{1}^{(1)}$ are given by: \[\Bigg\{e_{1}=
\begin{pmatrix}
  0 & 1 \\
  0 & 0
   \end{pmatrix}=e,
f_{1}=  \begin{pmatrix}
   0 & 0 \\
  1 & 0 
   \end{pmatrix}=f,
 e_{2}=  \begin{pmatrix}
   0 & 0 \\
  t & 0
   \end{pmatrix}=tf,
f_{2}=  \begin{pmatrix}
   0 & t^{-1} \\
  0 & 0
   \end{pmatrix}=t^{-1}e,
h=  \begin{pmatrix}
   1 & 0 \\
  0 & -1 
   \end{pmatrix}\bigg \}.\]
 The Cartan involution of $A_{1}^{(1)}$ is the following
 \begin{table}[h]
\centering
\begin{tabular}{c  c}
$t^{n}e\mapsto -t^{-n}f$ \quad \quad & $it^{n}e\mapsto it^{-n}f$\\
$t^{n}f\mapsto -t^{-n}e$ \quad \quad & $it^{n}f\mapsto it^{-n}e$\\
$t^{n}h\mapsto -t^{-n}h$ \quad \quad & $it^{n}h\mapsto it^{-n}h$\\
$c\mapsto -c$ & $ic\mapsto ic$\\
$d\mapsto -d$ & $id\mapsto id$\\
\end{tabular}
\end{table}
\newline 
As a result the compact form is generated by:
$\{ (t^{n}e-t^{-n}f), i(t^{n}e+t^{-n}f), (t^{n}f-t^{-n}e), i(t^{n}f+t^{-n}e), (t^{n}h-t^{-n}h), i(t^{n}h+ t^{-n}h)\vert n \in \mathbb{Z}\} \oplus \mathbb{R}i c \oplus \mathbb{R}id$.
Explicitly the compact real form is given by
$
\mathcal{C}(t) =\mathop{\hbox{\large$\sum$}}\limits_{n \in \mathbb{Z}}[a_{1}^{(n)}(t^{n}e-t^{-n}f)+i a_{2}^{(n)} (t^{n}e+t^{-n}f)+a_{3}^{(n)}
(t^{n}f-t^{-n}e)+ ia_{4}^{(n)}(t^{n}f +t^{-n}e)+a_{5}^{(n)}
(t^{n}h-t^{-n}h)+ia_{6}^{(n)} (t^{n}h+ t^{-n}h)] \oplus \mathbb{R}ic \oplus \mathbb{R}id $
which is equal to the following matrix\\ \\
$\mathcal{C}(t)= \mathop{\hbox{\Huge$\sum$}}\limits_{n \in \mathbb{Z}}\begin{pmatrix}
a_{5}^{(n)}(t^{n}-t^{-n})+ ia_{6}^{(n)}(t^{n}+t^{-n}) & t^{n}(a_{1}^{(n)}+ia_{2}^{(n)})+t^{-n}(-a_{3}^{(n)}+ia_{4}^{(n)}) \\
  t^{-n}(-a_{1}^{(n)}+ia_{2}^{(n)})+t^{n}(a_{3}^{(n)}+ia_{4}^{(n)}) &
  a_{5}^{(n)}(t^{-n}-t^{n})-ia_{6}^{(n)}(t^{n}+t^{-n})
\end{pmatrix}\oplus \mathbb{R}ic \begin{pmatrix}
   1 & 0 \\
  0 & 1 
   \end{pmatrix} \oplus \mathbb{R}id \begin{pmatrix}
   1 & 0 \\
  0 & 1 
   \end{pmatrix} 
$\\
this is a skew hermitian matrix with trace zero, which is identified as
$su^{(1)}(2)$. The Vogan diagram associated with this real form is given by the following:
\begin{figure}[h]
\centering
$\xy
\POS (40,0) *\cir<3pt>{}="a",
\POS (60,0) *\cir<3pt>{} ="b",
\POS "b" \ar@{<=>}^<<<{\alpha_{1}} "a",
\POS "a" \ar@{-}^<<<<{\alpha_{0}},
\endxy$ 
 \end{figure}
 \newline
Now the general form \cite{Cornwell1992} of an involutive automorphism associated with a affine Kac-Moody algebra is given as:
for type 1(a) automorphism
\begin{equation}\label{eq13}
\sigma(\mathcal{C}(t))= U(t) \mathcal{C}(u t)
U(t)^{-1}+\frac{1}{\gamma} Res\Bigg\{tr\Bigg(U(t)^{-1}\frac{dU(t)}{dt} \mathcal{C}(u(t))\Bigg)\Bigg\}c.
\end{equation}
and for type 1(b) automorphism
\begin{equation}
\sigma(\mathcal{C}(t))= U(t) (-\tilde{\mathcal{C}}(u t))
U(t)^{-1}+\frac{1}{\gamma} Res\Bigg\{tr\Bigg(U(t)^{-1}\frac{d U(t)}{d t}(-\tilde{\mathcal{C}}(u t))\Bigg)\Bigg\}c.
\end{equation}
But the conjugacy class of type 1(b) automorphisms with $u=1$ and $u= -1$ correspond to some automorphisms of type 1(a) with $u=1$ and $u= -1$ respectively. Also we shall like to mention that type 2(a) and type 2(b) automorphisms are obtained by composing type 1(a) and 1(b) with Cartan involution respectively.
  Action of $\sigma$ on $c$ is $\sigma(c)= \mu c$, however for 1(a) automorphism $\mu=1$. Now $\sigma(d)= \mu \Phi(U(t))+\xi c+ \mu d$ where $\Phi(U(t))$ is the $d_{\Gamma}\times d_{\Gamma}$ matrix that depends upon $U(t)$ as below, 
  \begin{equation*}
  \Phi(U(t))=\Bigg\{-t\frac{dU(t)}{dt}U(t)^{-1}+\frac{1}{d_{\Gamma}}tr\Bigg(t\frac{dU(t)}{dt}U(t)^{-1}\Bigg)I\Bigg\}
  \end{equation*}
  and for our cases $\sigma(d)=d$ except the case-III of $A_{1}^{(1)}$ where\\ $\sigma(d)=\begin{pmatrix}
   t^{2}/2 & 0 \\
  0 & -t^{2}/2
   \end{pmatrix}+d$. \\
\textbf{Case-I:} 
If
\begin{equation} \label{eq14}
 U(t)=U(t)^{-1}=  \begin{pmatrix}
   1 & 0 \\
  0 & -1
   \end{pmatrix},
  u=1,\quad \xi=0,
  \end{equation} then
under the automorphism (\ref{eq13}) with (\ref{eq14}) a general matrix with block matrices $A, B, C, D$ transforms as
\begin{equation*}\label{eq24}
 \begin{pmatrix}
 A & B  \\
 C & D
 \end{pmatrix} \longrightarrow \begin{pmatrix}
 A & -B  \\
 -C & D
 \end{pmatrix}.
 \end{equation*}
Hence the fixed subalgebra $K$ of
$\mathcal{C}(t)$ is given by
\begin{equation*}\label{eq15}
K=\mathop{\hbox{\Huge$\sum$}}\limits_{n \in \mathbb{Z}}\begin{pmatrix}
  a_{5}^{(n)}(t^{n}-t^{-n})+ ia_{6}^{(n)}(t^{n}+t^{-n}) & 0 \\
  0 & a_{5}^{(n)}(t^{-n}-t^{n})-ia_{6}^{(n)}(t^{n}+t^{-n})
\end{pmatrix}
\end{equation*}and
\begin{equation*}\label{eq16}
P=\mathop{\hbox{\Huge$\sum$}}\limits_{n \in \mathbb{Z}}\begin{pmatrix}
   0 & t^{n}(a_{1}^{(n)}+ia_{2}^{(n)})+t^{-n}(-a_{3}^{(n)}+ia_{4}^{(n)}) \\
 t^{-n}(-a_{1}^{(n)}+ia_{2}^{(n)})+t^{n}(a_{3}^{(n)}+ia_{4}^{(n)}) & 0 
\end{pmatrix}.
\end{equation*}
Thus
$
K+iP=\mathop{\hbox{\Huge$\sum$}}\limits_{n \in \mathbb{Z}}\begin{pmatrix} a_{5}^{(n)}(t^{n}-t^{-n})+ia_{6}^{((n)}(t^{n}+t^{-n}) &
t^{n}(-a_{2}^{(n)}+ia_{1}^{(n)})+t^{-n}(-ia_{3}^{(n)}-a_{4}^{(n)}) \\
 t^{-n}(-ia_{1}^{(n)}-a_{2}^{(n)})+t^{n}(ia_{3}^{(n)}-a_{4}^{(n)}) &
 a_{5}^{(n)}(t^{-n}-t^{n})-ia_{6}^{(n)}(t^{n}+t^{-n}).
 \end{pmatrix}$
Hence the non-compact real form is   $K+iP\oplus\mathbb{R}ic\oplus\mathbb{R}id \in su_{1}^{(1)}(1, 1)$ and the corresponding Vogan diagram is given by 
\begin{figure}[h]
\centering
$\xy
\POS (40,0) *{\bullet}="a",
\POS (60,0) *{\bullet} ="b",
\POS "b" \ar@{<=>}^<<<{\alpha_{1}} "a",
\POS "a" \ar@{-}^<<<<{\alpha_{0}},
\endxy$ 
 \end{figure}
 \newline
 and the two affine Kac-Moody symmetric spaces are
\begin{equation}\label{eq17}
SU_{1}^{(1)}(1, 1)/S_{1}^{(1)}(U_{1}\times U_{1}), \quad
SU^{(1)}(1+1)/S_{1}^{(1)}(U_{1}\times U_{1}).
\end{equation}
\textbf{Case II:}
Similarly if
\begin{equation}\label{eq18}
 U(t)=U(t)^{-1}=  \begin{pmatrix}
   1 & 0 \\
  0 & -1 
   \end{pmatrix},\quad
  u=-1,\quad \xi=0
  \end{equation}
then under the automorphism(\ref{eq13}) with (\ref{eq18}) the matrix $\begin{pmatrix}
 A & B  \\
 C & D
 \end{pmatrix}$ trasforms as \begin{equation*}\label{eq24}
 \begin{pmatrix}
 A & B  \\
 C & D
 \end{pmatrix} \longrightarrow \begin{pmatrix}
 -A & B  \\
 C & -D
 \end{pmatrix}.
 \end{equation*}
We observe when $n$ is an even integer it reduces to case I, giving same real form. But $n$ is odd then we have 
  \begin{equation*}\label{eq19}
  K= \mathop{\hbox{\Huge$\sum$}}\limits_{n \in \mathbb{Z}}\begin{pmatrix}
 0 & t^{n}(a_{1}^{(n)}+ia_{2}^{(n)})+t^{-n}(-a_{3}^{(n)}+ia_{4}^{(n)}) \\
 t^{-n}(-a_{1}^{(n)}+ia_{2}^{(n)})+t^{n}(a_{3}^{(n)}+ia_{4}^{(n)}) & 0
\end{pmatrix} \end{equation*} and
\begin{equation*}\label{eq20}
P= \mathop{\hbox{\Huge$\sum$}}\limits_{n \in \mathbb{Z}}\begin{pmatrix} a_{5}^{(n)}(t^{n}-t^{-n})+ia_{6}^{(n)}(t^{n}+t^{-n}) & 0 \\
0   & t^{-n}(-a_{1}^{(n)}+ ia_{2}^{(n)})+t^{n}(a_{3}^{(n)}+ia_{4}^{(n)})
\end{pmatrix}.
\end{equation*}

$ K+iP= \mathop{\hbox{\Huge$\sum$}}\limits_{n \in \mathbb{Z}} \begin{pmatrix}
ia_{5}^{(n)}(t^{n}-t^{-n})-a_{6}^{(n)}(t^{n}+t^{-n}) &
t^{n}(a_{1}^{(n)}+ia_{2}^{(n)})+t^{-n}(-a_{3}^{(n)}+ia_{4}^{(n)})  \\
 t^{-n}(-a_{1}^{(n)}+ia_{2}^{(n)})+t^{n}(a_{3}^{(n)}+ia_{4}^{(n)}) &
 ia_{5}^{(n)}(t^{-n}-t^{n})+a_{6}^{(n)}(t^{n}+t^{-n}).
\end{pmatrix}$
Hence the non-compact real form is   $K+iP\oplus\mathbb{R}ic\oplus\mathbb{R}id \in su_{-1}^{(1)}(1, 1)$. The corresponding Vogan diagram is the given by
\begin{figure}[h]
\centering
$\xy
\POS (40,0) *\cir<3pt>{}="a",
\POS (60,0) *{\bullet} ="b",
\POS "b" \ar@{<=>}^<<<{\alpha_{1}} "a",
\POS "a" \ar@{-}^<<<<{\alpha_{0}},
\endxy$ 
 \end{figure}
 \newline
The affine Kac-Moody symmetric spaces are:
\begin{equation}\label{eq22}
SU_{-1}^{(1)}(1, 1)/S_{-1}^{(1)}(U_{1}\times U_{1}), \quad
SU^{(1)}(1+1)/S_{-1}^{(1)}(U_{1}\times U_{1}).
\end{equation}
\textbf{Case III:}
Now consider
\begin{equation}\label{eq23}
U(t)= \begin{pmatrix}
 0 & 1  \\
 -t & 0
 \end{pmatrix}, \quad u=1, \quad \xi=-1, \quad U(t)^{-1}= \begin{pmatrix}
 0 & -t  \\
 1 & 0
 \end{pmatrix}.
 \end{equation}
 So now under the automorphism (\ref{eq13}) with (\ref{eq23}) a general matrix $\begin{pmatrix}
 A & B  \\
 C & D
 \end{pmatrix}$ transform as,
 \begin{equation}\label{eq24}
 \begin{pmatrix}
 A & B  \\
 C & D
 \end{pmatrix} \longrightarrow \begin{pmatrix}
 D & -Ct^{-1}  \\
 -Bt & A
 \end{pmatrix}.
 \end{equation}
Then from a simple mathematical manipulation we observe that in this case
 \begin{equation}\label{eq25}
 K=\begin{pmatrix}
 A+D & B-Ct^{-1}  \\
 C-Bt & D+A
 \end{pmatrix},\quad P= \begin{pmatrix}
 A-D & B+Ct^{-1}  \\
 C+Bt & D-A.
 \end{pmatrix}
 \end{equation}
 Now putting the values of A, B, C and D in $K$ we get

 $K=\\\frac{1}{2}\mathop{\hbox{\Huge$\sum$}}\limits_{n \in \mathbb{Z}} \begin{pmatrix}
 0 & t^{n}(a_{1}^{(n)}+ia_{2}^{(n)})+t^{-n}(-a_{3}^{(n)}+ia_{4}^{(n)})+\\ &t^{-(n+1)}(a_{1}^{(n)}-ia_{2}^{(n)})-(a_{3}^{(n)}+ia_{4}^{(n)})t^{n-1}  \\
 t^{-n}(-a_{1}^{(n)}+ia_{2}^{(n)})+t^{n}(a_{3}^{(n)}+ia_{4}^{(n)})\cr+t^{n+1}(-a_{1}^{(n)}-ia_{2}^{(n)})+(a_{3}^{(n)}-ia_{4}^{(n)})t^{-n+1}  &
 0
 \end{pmatrix}$

with $K$ satisfies $K^{*}+K=0$ and hence $K $ is identified as $so^{(1)}(2)$. So we have  $ K+iP= \\
\frac{1}{2}\mathop{\hbox{\Huge$\sum$}}\limits_{n \in \mathbb{Z}}\begin{pmatrix}
a_{5}^{(n)}(t^{n}-t^{-n})+ia_{6}^{(n)}(t^{n}+t^{-n}) &
t^{n}(a_{1}^{(n)}+ia_{2}^{(n)})+t^{-n}(-a_{3}^{(n)}+ia_{4}^{(n)})+\\ &
t^{-(n+1)}(a_{1}^{(n)}-ia_{2}^{(n)})-(a_{3}^{(n)}+ia_{4}^{(n)})t^{n-1}  \\
 t^{-n}(-a_{1}^{(n)}+ia_{2}^{(n)})+t^{n}(a_{3}^{(n)}+ia_{4}^{(n)})+\\ t^{n+1}(-a_{1}^{(n)}-ia_{2}^{(n)})+(a_{3}^{(n)}-ia_{4}^{(n)})t^{-n+1}  &
 a_{5}^{(n)}(-t^{n}+t^{-n})-ia_{6}^{(n)}(t^{n}+t^{-n})
 \end{pmatrix} \oplus \mathbb{R}ic\begin{pmatrix}
 1 & 0  \\
 0 & 1
 \end{pmatrix} \oplus \mathbb{R}id\begin{pmatrix}
 1 & 0  \\
 0 & 1
 \end{pmatrix}$ which is identified as $sl^{(1)}(2, \mathbb{R})$ and the Vogan diagram is 
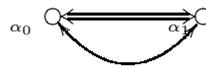
\begin{figure}[h]
\centering
$\xy
\POS (40,0) *\cir<3pt>{}="a",
\POS (60,0) *\cir<3pt>{}="b",
\POS "b" \ar@{<=>}^<<<{\alpha_{1}} "a",
\POS "a" \ar@{-}^<<<<{\alpha_{0}},
\POS "a" \ar@{<->}@/_{1.5pc}/ "b",
\endxy$ 
\end{figure}
\newline
 Hence the affine Kac-Moody symmetric spaces are
\begin{equation}\label{eq28}
SU^{(1)}(2)/SO^{(1)}(2), \quad SL^{(1)}(2,\mathbb{R})/SO^{(1)}(2).
\end{equation} 
\subsection{Affine Kac Moody symetric space associated with $A_{2}^{(1)}$}
The
Chevelley generators of $A_{2}^{(1)}$ are:
$\Bigg\{e_{1}=
\begin{pmatrix}
 0 & 1 & 0 \\
  0 & 0 & 0 \\
  0 & 0 & 0 
   \end{pmatrix},
f_{1}=\begin{pmatrix}
  0 & 0 & 0 \\
  1 & 0 & 0 \\
  0 & 0 & 0 
\end{pmatrix},
 h_{1}=\begin{pmatrix}
   1 & 0 & 0\\
  0 & -1 & 0 \\
  0 & 0 & 0 
   \end{pmatrix},
   e_{2}=  \begin{pmatrix}
  0 & 0 & 0\\
  0 & 0 & 1\\
  0 & 0 & 0
   \end{pmatrix},
f_{2}=  \begin{pmatrix}
   0 & 0 & 0 \\
   0 & 0 & 0\\
   0 & 1 & 0
   \end{pmatrix},
h_{2}=  \begin{pmatrix}
   0 & 0 & 0 \\
   0 & 1 & 0 \\
   0 & 0 & -1 
   \end{pmatrix},
   e_{3}=  \begin{pmatrix}
  0 & 0 & 0\\
  0 & 0 & 0 \\
  t & 0 & 0 
   \end{pmatrix},
f_{3}=  \begin{pmatrix}
   0 & 0 & t^{-1} \\
   0 & 0 & 0 \\
   0 & 0 & 0
   \end{pmatrix}
h_{3}=  \begin{pmatrix}
   -1 & 0 & 0 \\
   0 & 0 & 0 \\
   0 & 0 & 1
   \end{pmatrix}\Bigg\}.$
The Cartan involution of $A_{2}^{(1)}$ is the following:

\begin{table}[h]
\centering
\begin{tabular}{c  c}
$t^{n}e_{1}\mapsto -t^{-n}f_{1}$ \quad \quad & $it^{n}e_{1}\mapsto it^{-n}f_{1}$\\
$t^{n}e_{2}\mapsto -t^{-n}f_{2}$ \quad \quad & $it^{n}e_{2}\mapsto it^{-n}f_{2}$\\
$t^{n}e_{3}\mapsto -t^{-n}f_{3}$ \quad \quad & $it^{n}e_{3}\mapsto it^{-n}f_{3}$\\
$t^{n}f_{1}\mapsto -t^{-n}e_{1}$ \quad \quad & $it^{n}f_{1}\mapsto it^{-n}e_{1}$\\
$t^{n}f_{2}\mapsto -t^{-n}e_{2}$ \quad \quad & $it^{n}f_{2}\mapsto it^{-n}e_{2}$\\
$t^{n}f_{3}\mapsto -t^{-n}e_{3}$ \quad \quad & $it^{n}f_{3}\mapsto it^{-n}e_{3}$\\
$t^{n}h_{1}\mapsto -t^{-n}h_{1}$ \quad \quad & $it^{n}h_{1}\mapsto it^{-n}h_{1}$\\
$t^{n}h_{2}\mapsto -t^{-n}h_{2}$ \quad \quad & $it^{n}h_{2}\mapsto it^{-n}h_{2}$\\
$t^{n}h_{3}\mapsto -t^{-n}h_{3}$ \quad \quad & $it^{n}h_{3}\mapsto it^{-n}h_{3}$\\
$c\mapsto -c$ & $ic\mapsto ic$\\
$d\mapsto -d$ & $id\mapsto id$\\
\end{tabular}
\end{table} 
Hence the Compact form is generated
by $\{ e_{1}t^{n}-f_{1}t^{-n}, i(e_{1}t^{n}+f_{1}t^{-n}),
h_{1}t^{n}-h_{1}t^{-n}, i(h_{1}t^{n}+h_{1}t^{-n}),
e_{2}t^{n}-f_{2}t^{-n}, i(e_{2}t^{n}+f_{2}t^{-n}),
h_{2}t^{n}-h_{2}t^{-n}, i(h_{2}t^{n}+h_{2}t^{-n}),
e_{3}t^{n}-f_{3}t^{-n}, i(e_{3}t^{n}+f_{3}t^{-n}),
h_{3}t^{n}-h_{3}t^{-n}, i(h_{3}t^{n}+h_{3}t^{-n}\mid n \in
\mathbb{Z}\} \oplus \mathbb{R}ic \oplus \mathbb{R}id$.
 So, the compact form $\mathcal{C}(t)$ is
$ \mathop{\hbox{\Huge$\sum$}}\limits_{n \in \mathbb{Z}}\begin{pmatrix}
   A_{11}^{(n)} & A_{12}^{(n)} & A_{13}^{(n)} \\
   A_{21}^{(n)} & A_{22}^{(n)} & A_{23}^{(n)} \\
    A_{31}^{(n)} & A_{32}^{(n)} & A_{33}^{(n)} 
   \end{pmatrix} \oplus \mathbb{R}ic \begin{pmatrix}
   1 & 0 & 0\\
  0 & 1 & 0 \\
  0 & 0 & 1
   \end{pmatrix} \oplus \mathbb{R}id \begin{pmatrix}
   1 & 0 & 0\\
  0 & 1 & 0 \\
  0 & 0 & 1 
   \end{pmatrix}$

where
\begin{eqnarray*}\label{a4}
A_{11}^{(n)}&=&a_{3}^{(n)}(t^{n}-t^{-n})+ia_{4}^{(n)}(t^{n}+t^{-n})-a_{11}^{(n)}(t^{n}-t^{-n})-ia_{12}^{(n)}(t^{n}+t^{-n})\\
A_{12}^{(n)}&=&(a_{1}^{(n)}+ia_{2}^{(n)})t^{n}\\
A_{13}^{(n)}&=&-a_{9}^{(n)}t^{-(n+1)}+ ia_{10}^{(n)}t^{-(n+1)}\\
A_{21}^{(n)}&=&(-a_{1}^{(n)}+ ia_{2}^{(n)})t^{-n}\\
A_{22}^{(n)}&=&a_{3}^{(n)}(-t^{n}+t^{-n})-ia_{4}^{(n)}(t^{n}+t^{-n})+a_{7}^{(n)}(t^{n}-t^{-n})+ia_{8}^{(n)}(t^{n}+t^{-n})\\
A_{23}^{(n)}&=&(a_{5}^{(n)}+ ia_{6}^{(n)})t^{n}\\
A_{31}^{(n)}&=&(ia_{10}^{(n)}+a_{9}^{(n)})t^{n+1}\\
A_{32}^{(n)}&=&(-a_{5}^{(n)}+ia_{6}^{(n)})t^{-n} \\
A_{33}^{(n)}&=&a_{7}^{(n)}(-t^{n}+t^{-n})-ia_{8}^{(n)}(t^{n}+t^{-n})
+a_{11}^{(n)}(t^{n}-t^{-n})+ ia_{12}^{(n)}(t^{n}+t^{-n})
\end{eqnarray*}

This matrix is in the form:
\begin{equation}\label{eq40}
\begin{pmatrix}
(A)_{2\times 2} & (B)_{2\times 1}\\ (-B^{*})_{1\times 2} &
(C)_{1\times 1}
\end{pmatrix},
\end{equation}
which is a skew hermitian matrix with trace zero it is identified as $su^{(1)}(3)$. The Vogan diagram is 
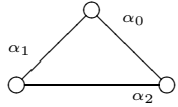
\begin{figure}[h]
\centering
$\xy
\POS (40,0) *\cir<3pt>{}="a",
\POS (60,0) *\cir<3pt>{}="b",
\POS (50,10) *\cir<3pt>{}="c"
\POS "a" \ar@{-}^<<<<{\alpha_{1}}, "c",
\POS "b" \ar@{-}^<<<{\alpha_{2}} "a",
\POS "c" \ar@{-}^<<<<{\alpha_{0}}, "b",

\endxy$
\end{figure} 
\newline
 Now proceeding similarly as $A_{1}^{(1)}$ case and taking different cases we have,\\\\
\textbf{Case I:}
\begin{equation}\label{eq41}
U(t)=U(t)^{-1}=\begin{pmatrix} 1 & 0 & 0 \\
 0 & 1 & 0\\
 0 & 0 & -1
 \end{pmatrix}, \quad u=1, \quad \xi=0 .
 \end{equation}
 Under the automorphism (\ref{eq13}) with (\ref{eq41}) the matrix transforms as
 \begin{equation}\label{eq27}
 \begin{pmatrix}
 A & B \\
 (-B^{*}) & C
 \end{pmatrix} \longrightarrow \begin{pmatrix}
 A & -B \\
 B^{*} & C
 \end{pmatrix}
 \end{equation}
 Here \begin{equation}\label{eq42}
 K=\begin{pmatrix}
 A & 0 \\
 0& C
 \end{pmatrix},\quad P= \begin{pmatrix}
0& B \\
 -B^{*} & 0.
 \end{pmatrix}
 \end{equation}
 The decomposition of $K$ as in (\ref{eq37})
 shows that is isomorphic to $su_{1}^{(1)}(2)\times c_{0}\times su_{1}^{(1)}(1)$ where $c_{0}$ is the center of $K$. Now  $K+iP$ is
\begin{equation}\label{eq44}
\begin{pmatrix}
(Z_{1})_{2\times 2} & (Z_{2})_{2\times 1}\\ (Z_{2}^{*})_{1\times
2} & (Z_{3})_{1\times 1}
\end{pmatrix}
\end{equation}
with $Z_{2}=iB$ is a $2\times 2$ matrix, $Z_{1}=A$ is a $2\times 2$
skew hermitian matrix and $Z_{3}=C$ is a $1\times 1$ skew hermitian
matrix and also satisfies $Tr Z_{1}+Tr Z_{3}=0$. Hence the non-compact real form is   $K+iP\oplus\mathbb{R}ic\oplus\mathbb{R}id \in
su_{1}^{(1)}(2,1)$. The Vogan diagram is
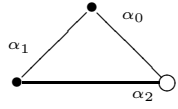
\begin{figure}[h]
\centering
$\xy
\POS (40,0) *{\bullet}="a",
\POS (60,0) *\cir<3pt>{}="b",
\POS (50,10) *{\bullet}="c"
\POS "a" \ar@{-}^<<<<{\alpha_{1}}, "c",
\POS "b" \ar@{-}^<<<{\alpha_{2}} "a",
\POS "c" \ar@{-}^<<<<{\alpha_{0}}, "b",
\endxy$
\end{figure} 
\newline
 Thus the corresponding symmetric
spaces are:
\begin{equation}\label{eq45}
SU_{1}^{(1)}(2,1)/S_{1}^{(1)}(U_{2}\times U_{1}),\quad
SU^{(1)}(2+1)/S_{1}^{(1)}(U_{2}\times U_{1}).
\end{equation}
\textbf{Case II:}
\begin{equation}\label{eq46}
U(t)=U(t)^{-1}=\begin{pmatrix}
 1 & 0 & 0 \\
 0 & 1 & 0\\
 0 & 0 & -1
 \end{pmatrix}, \quad u=-1,\quad \xi=0.
 \end{equation}
 Under the automorphism (\ref{eq13}) with (\ref{eq46}) the matrix transforms as
 \begin{equation}\label{eq47}
 \begin{pmatrix}
 A & B \\
 -B^{*} & C
 \end{pmatrix} \longrightarrow \begin{pmatrix}
 -A & B \\
 -B^{*} & -C
 \end{pmatrix}.
 \end{equation}
 Here \begin{equation}\label{eq48}
 K=\begin{pmatrix}
 0 & B \\
-B^{*}& 0
 \end{pmatrix},\quad P= \begin{pmatrix}
A& 0 \\
 0 & C
 \end{pmatrix}
 \end{equation}
Thus $K+iP$ is
\begin{equation}\label{eq49}
\begin{pmatrix}
(Z_{1})_{2\times 2} & (Z_{2})_{2\times 1}\\ (-Z_{2}^{*})_{1\times
2} & (Z_{3})_{1\times 1}
\end{pmatrix}
\end{equation}
with $Z_{2}=iB$ is a $2\times 1$ matrix, $Z_{1}=A$ is a $2\times 2$
skew hermitian matrix and $Z_{3}=C$ is a $1\times 1$ skew hermitian
matrix and also satisfies $Tr Z_{1}+Tr Z_{3}=0$. Hence the non-compact real form is   $K+iP\oplus\mathbb{R}ic\oplus\mathbb{R}id \in 
su_{-1}^{(1)}(2,1)$. The Vogan diagram is
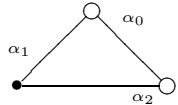
\begin{figure}[h]
\centering
$\xy
\POS (40,0) *{\bullet}="a",
\POS (60,0) *\cir<3pt>{}="b",
\POS (50,10) *\cir<3pt>{}="c"
\POS "a" \ar@{-}^<<<<{\alpha_{1}}, "c",
\POS "b" \ar@{-}^<<<{\alpha_{2}} "a",
\POS "c" \ar@{-}^<<<<{\alpha_{0}}, "b",
\endxy$
\end{figure}
\newline
 Thus the corresponding
symmetric spaces are:
\begin{equation}\label{eq50}
SU_{-1}^{(1)}(2,1)/S_{-1}^{(1)}(U_{2}\times U_{1}),\quad
SU^{(1)}(2+1)/S_{-1}^{(1)}(U_{2}\times U_{1}).
\end{equation}
\textbf{Case III:}
\begin{equation}\label{eq52}
U(t)=\begin{pmatrix}
 0 & 0 & 1 \\
 0 & 1 & 0\\
 1 & 0 & 0
 \end{pmatrix}, \quad u=1,\quad \xi=0, \quad  U(t)^{-1}=\begin{pmatrix}
 0 & 0 & -1 \\
 0 & -1 & 0\\
 -1 & 0 & 0
 \end{pmatrix}.
\end{equation}
 Under the automorphism (\ref{eq13}) with (\ref{eq52}) the matrix transforms as
 \begin{equation}\label{eq53}
\mathop{\hbox{\Huge$\sum$}}\limits_{n \in \mathbb{Z}} \begin{pmatrix}
 A_{11}^{(n)} & A_{12}^{(n)} & A_{13}^{(n)} \\
 A_{21}^{(n)} & A_{22}^{(n)} & A_{23}^{(n)} \\
 A_{31}^{(n)} & A_{32}^{(n)} & A_{33}^{(n)} 
\end{pmatrix} \longrightarrow \mathop{\hbox{\Huge$\sum$}}\limits_{n \in \mathbb{Z}}\begin{pmatrix}
  A_{33}^{(n)} & A_{32}^{(n)} & A_{31}^{(n)} \\
 A_{23}^{(n)} & A_{22}^{(n)} & A_{21}^{(n)} \\
 A_{13}^{(n)} & A_{12}^{(n)} & A_{11}^{(n)} 
 \end{pmatrix}.
 \end{equation}
$K=\\\frac{1}{2}\mathop{\hbox{\Huge$\sum$}}\limits_{n \in \mathbb{Z}}\begin{pmatrix}
 A_{11}^{(n)}+A_{33}^{(n)} & A_{12}^{(n)}+A_{32}^{(n)} & A_{13}^{(n)}+A_{31}^{(n)} \\
 A_{21}^{(n)}+A_{23}^{(n)} & A_{22}^{(n)} & A_{23}^{(n)}+A_{21}^{(n)} \\
 A_{31}^{(n)}+ A_{13}^{(n)} & A_{32}^{(n)}+A_{12}^{(n)}  & A_{33}^{(n)}+A_{11}^{(n)} 
\end{pmatrix}=\frac{1}{2}\mathop{\hbox{\Huge$\sum$}}\limits_{n \in \mathbb{Z}}\begin{pmatrix}
 (\widehat{A}_{11}^{(n)})_{+} & (\widehat{A}_{12}^{(n)})_{+} & (\widehat{A}_{13}^{(n)})_{+} \\
 (\widehat{A}_{21}^{(n)})_{+} & (\widehat{A}_{22}^{(n)})_{+} & (\widehat{A}_{23}^{(n)})_{+}\\
 (\widehat{A}_{31}^{(n)})_{+} & (\widehat{A}_{32}^{(n)})_{+} & (\widehat{A}_{33}^{(n)})_{+}
 
\end{pmatrix}$\\ and
 $P=\\ \frac{1}{2}\mathop{\hbox{\Huge$\sum$}}\limits_{n \in \mathbb{Z}}\begin{pmatrix}
 A_{11}^{(n)}-A_{33}^{(n)} & A_{12}^{(n)}-A_{32}^{(n)} & A_{13}^{(n)}-A_{31}^{(n)} \\
 A_{21}^{(n)}-A_{23}^{(n)}& 0 & A_{23}^{(n)}-A_{21}^{(n)} \\
 A_{31}^{(n)}-A_{13}^{(n)} & A_{32}^{(n)}-A_{12}^{(n)} & A_{33}^{(n)}-A_{11}^{(n)} 
\end{pmatrix}=\frac{1}{2}\mathop{\hbox{\Huge$\sum$}}\limits_{n \in \mathbb{Z}}\begin{pmatrix}
 (\widehat{A}_{11}^{(n)})_{-} & (\widehat{A}_{12}^{(n)})_{-} & (\widehat{A}_{13}^{(n)})_{-}\\
 (\widehat{A}_{21}^{(n)})_{-} & 0 & (\widehat{A}_{23}^{(n)})_{-}\\
 (\widehat{A}_{31}^{(n)})_{-} & (\widehat{A}_{32}^{(n)})_{-} & (\widehat{A}_{33}^{(n)})_{-}
 
\end{pmatrix}$.
Thus $K+iP$ is
\begin{equation}\label{eq56}
\frac{1}{2}\mathop{\hbox{\Huge$\sum$}}\limits_{n \in \mathbb{Z}}\begin{pmatrix}
 (\widehat{A}_{11}^{(n)})_{+}+i(\widehat{A}_{11}^{(n)})_{-} & (\widehat{A}_{12}^{(n)})_{+}+i(\widehat{A}_{12}^{(n)})_{-} & (\widehat{A}_{13}^{(n)})_{+}+i(\widehat{A}_{13}^{(n)})_{-} \\
 (\widehat{A}_{21}^{(n)})_{+}+i(\widehat{A}_{21}^{(n)})_{-} & (\widehat{A}_{22}^{(n)})_{+} & (\widehat{A}_{23}^{(n)})_{+}+i(\widehat{A}_{23}^{(n)})_{-}\\
 (\widehat{A}_{31}^{(n)})_{+}+i(\widehat{A}_{31}^{(n)})_{-} & (\widehat{A}_{32}^{(n)})_{+}+i(\widehat{A}_{32}^{(n)})_{-} & (\widehat{A}_{33}^{(n)})_{+}+i(\widehat{A}_{33}^{(n)})_{-}
 
\end{pmatrix}
\end{equation}
such that trace of this matrix is zero. Hence the non-compact real form is   $K+iP\oplus\mathbb{R}ic\oplus\mathbb{R}id \in sl_{1}^{(1)}(3, \mathbb{R})$. Here $K=
so_{1}^{(1)}(3)$. The Vogan diagram is
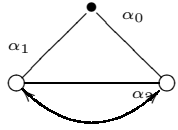
\begin{figure}[h]
\centering
$\xy
\POS (40,0) *\cir<3pt>{}="a",
\POS (60,0) *\cir<3pt>{}="b",
\POS (50,10) *{\bullet}="c"
\POS "a" \ar@{-}^<<<<{\alpha_{1}}, "c",
\POS "b" \ar@{-}^<<<{\alpha_{2}} "a",
\POS "c" \ar@{-}^<<<<{\alpha_{0}}, "b",
\POS "a" \ar@{<->}@/_{1.25pc}/ "b",
\endxy$
\end{figure} 
\newline
Thus the corresponding symmetric spaces are
\begin{equation}\label{eq57}
SL_{1}^{(1)}(3,\mathbb{R})/SO_{1}^{(1)}(3), \quad
SU^{(1)}(3)/SO_{1}^{(1)}(3).
\end{equation}
\textbf{Case IV:}
\begin{equation}\label{eq58}
U(t)=\begin{pmatrix}
 0 & 0 & 1 \\
 0 & 1 & 0\\
 1 & 0 & 0
 \end{pmatrix}, \quad u=-1,\quad \xi=0, \quad  U(t)^{-1}=\begin{pmatrix}
 0 & 0 & -1 \\
 0 & -1 & 0\\
 -1 & 0 & 0
 \end{pmatrix}
 \end{equation}
 Under the automorphism (\ref{eq13}) with (\ref{eq58}) the matrix transforms as
 \begin{equation}\label{eq59}
 \mathop{\hbox{\Huge$\sum$}}\limits_{n \in \mathbb{Z}}\begin{pmatrix}
 A_{11}^{(n)} & A_{12}^{(n)} & A_{13}^{(n)} \\
 A_{21}^{(n)} & A_{22}^{(n)} & A_{23}^{(n)} \\
 A_{31}^{(n)} & A_{32}^{(n)} & A_{33}^{(n)}
\end{pmatrix} \longrightarrow \mathop{\hbox{\Huge$\sum$}}\limits_{n \in \mathbb{Z}}\begin{pmatrix}
 -A_{33}^{(n)} & -A_{32}^{(n)} & -A_{31}^{(n)} \\
 -A_{23}^{(n)} & -A_{22}^{(n)} & -A_{21}^{(n)} \\
 -A_{13}^{(n)} & -A_{12}^{(n)} & -A_{11}^{(n)}
 \end{pmatrix}.
 \end{equation}
 Here
$ K=\\
\frac{1}{2}\mathop{\hbox{\Huge$\sum$}}\limits_{n \in \mathbb{Z}}\begin{pmatrix}
 A_{11}^{(n)}-A_{33}^{(n)} & A_{12}^{(n)}-A_{32}^{(n)} & A_{13}^{(n)}-A_{31}^{(n)} \\
 A_{21}^{(n)}-A_{23}^{(n)} & 0 & A_{23}^{(n)}-A_{21}^{(n)} \\
 A_{31}^{(n)}- A_{13}^{(n)} & A_{32}^{(n)}-A_{12}^{(n)} & A_{33}^{(n)}-A_{11}^{(n)} 
\end{pmatrix}=\frac{1}{2}\mathop{\hbox{\Huge$\sum$}}\limits_{n \in \mathbb{Z}}\begin{pmatrix}
 (\widehat{A}_{11}^{(n)})_{-} & (\widehat{A}_{12}^{(n)})_{-} & (\widehat{A}_{13}^{(n)})_{-} \\
 (\widehat{A}_{21}^{(n)})_{-} & 0 & (\widehat{A}_{23}^{(n)})_{-}\\
 (\widehat{A}_{31}^{(n)})_{-} & (\widehat{A}_{32}^{(n)})_{-} & (\widehat{A}_{33}^{(n)})_{-}
 
\end{pmatrix}$\\ and
 $P=\\
 \frac{1}{2}\mathop{\hbox{\Huge$\sum$}}\limits_{n \in \mathbb{Z}}\begin{pmatrix}
 A_{11}^{(n)}+A_{33}^{(n)} & A_{12}^{(n)}+A_{32}^{(n)} & A_{13}^{(n)}+A_{31}^{(n)} \\
 A_{21}^{(n)}+A_{23}^{(n)} & A_{22}^{(n)} & A_{23}^{(n)}+A_{21}^{(n)} \\
 A_{31}^{(n)}+ A_{13}^{(n)} & A_{32}^{(n)}+A_{12}^{(n)} & A_{11}^{(n)}+A_{33}^{(n)} 
\end{pmatrix}=\frac{1}{2}\mathop{\hbox{\Huge$\sum$}}\limits_{n \in \mathbb{Z}}\begin{pmatrix}
 (\widehat{A}_{11}^{(n)})_{+} & (\widehat{A}_{12}^{(n)})_{+} & (\widehat{A}_{13}^{(n)})_{+} \\
 (\widehat{A}_{21}^{(n)})_{+} & (\widehat{A}_{22}^{(n)})_{+} & (\widehat{A}_{23}^{(n)})_{+}\\
 (\widehat{A}_{31}^{(n)})_{+} & (\widehat{A}_{32}^{(n)})_{+} & (\widehat{A}_{33}^{(n)})_{+}
 
\end{pmatrix}$
Thus $K+iP$ is
\begin{equation}\label{eq62}
\frac{1}{2}\mathop{\hbox{\Huge$\sum$}}\limits_{n \in \mathbb{Z}}\begin{pmatrix}
 (\widehat{A}_{11}^{(n)})_{-}+i(\widehat{A}_{11}^{(n)})_{+} & (\widehat{A}_{12}^{(n)})_{-}+i(\widehat{A}_{12}^{(n)})_{+} & (\widehat{A}_{13}^{(n)})_{-}+i(\widehat{A}_{13}^{(n)})_{+} \\
 (\widehat{A}_{21}^{(n)})_{-}+i(\widehat{A}_{21}^{(n)})_{+} & i(\widehat{A}_{22}^{(n)})_{+} & (\widehat{A}_{23}^{(n)})_{-}+i(\widehat{A}_{23}^{(n)})_{+}\\
 (\widehat{A}_{31}^{(n)})_{-}+i(\widehat{A}_{31}^{(n)})_{+} & (\widehat{A}_{32}^{(n)})_{-}+i(\widehat{A}_{32}^{(n)})_{+} & (\widehat{A}_{33}^{(n)})_{-}+i(\widehat{A}_{33}^{(n)})_{+}
 
\end{pmatrix}
\end{equation}
such that trace of this matrix is zero. Hence the non-compact real form is   $K+iP\oplus\mathbb{R}ic\oplus\mathbb{R}id \in sl_{-1}^{(1)}(3, \mathbb{R})$. The Vogan diagram is
\begin{figure}[h]
\centering
$\xy
\POS (40,0) *\cir<3pt>{}="a",
\POS (60,0) *\cir<3pt>{}="b",
\POS (50,10) *\cir<3pt>{}="c"
\POS "a" \ar@{-}^<<<<{\alpha_{1}}, "c",
\POS "b" \ar@{-}^<<<{\alpha_{2}} "a",
\POS "c" \ar@{-}^<<<<{\alpha_{0}}, "b",
\POS "a" \ar@{<->}@/_{1.25pc}/ "b",

\endxy$
\end{figure}
\newline
 Here
$K= so_{-1}^{(1)}(3)$. Thus the corresponding symmetric spaces are
\begin{equation}\label{eq63}
SL_{-1}^{(1)}(3,\mathbb{R})/SO_{-1}^{(1)}(3), \quad
SU^{(1)}(3)/SO_{-1}^{(1)}(3).
\end{equation}
Thus we have completed the explicit (algebraically) calculation of affine Kac-Moody symmetric spaces associated with  $A_{1}^{(1)}$ and $A_{2}^{(1)}$ and relating them with the Vogan diagrams which classify their real forms also. The next tables contain the real forms, Vogan diagrams, fixed algebras associated with the automorphism, related compact and non-compact affine kac-Moody symmetric spaces. In order to have a clear idea about the computation of fixed algebras and affine Kac-Moody symmetric spaces we have  provided an appendix which contains all the necessary ingredients to understand the method of our calculations.
\begin{sidewaystable}[p]
\caption{Affine Kac-Moody symmetric spaces associated with $A_{2n-1}^{(1)}$}
\centering % centering table
\begin{tabular}{ | l| p{2cm} |l | p{2.5cm}| p{3cm} |p{3cm}|}
\hline
 Dynkin Diagram & Real Forms&Vogan Diagram &Fixed Algebra& Compact affine Kac-Moody Symmetric spaces&Non-compact affine Kac-Moody Symmetric spaces\\
 \hline
\multirow{8}{*}{$\xy
\POS (0,0) *\cir<3pt>{}="a",
\POS (10,0) *\cir<3pt>{} ="b",
\POS (20,0) *\cir<3pt>{} ="c",
\POS (30,0) *\cir<3pt>{} ="d",
\POS (40,0) *\cir<3pt>{} ="e",
\POS (50,0)  ="f"
\POS (20,10) *\cir<3pt>{} ="g",
\POS "a" \ar@{-}^<<<{\alpha_{1}} "b",
\POS "b" \ar@{.}^<<<{\alpha_{2}} "c",
\POS "c" \ar@{.}^<<<{\alpha_{n}} "d",
\POS "d" \ar@{-}^<<<{\alpha_{2n-2}} "e",
\POS "e" \ar@{}^<<<<{\alpha_{2n-1}} "f",
\POS "a" \ar@{-}^<<<<{} "g",
\POS "e" \ar@{-}^<<<<{} "g",
\POS "g" \ar@{}^<<{\alpha_{0}} "d",
\endxy $} & $\mathfrak{su}^{(1)}(2n)$& $\xy

\POS (0,0) *\cir<3pt>{}="a",
\POS (10,0) *\cir<3pt>{} ="b",
\POS (20,0) *\cir<3pt>{} ="c",
\POS (30,0) *\cir<3pt>{} ="d",
\POS (40,0) *\cir<3pt>{} ="e",
\POS (50,0)  ="f"
\POS (20,10) *\cir<3pt>{} ="g",
\POS "a" \ar@{-}^<<<{\alpha_{1}} "b",
\POS "b" \ar@{.}^<<<{\alpha_{2}} "c",
\POS "c" \ar@{.}^<<<{\alpha_{n}} "d",
\POS "d" \ar@{-}^<<<{\alpha_{2n-2}} "e",
\POS "e" \ar@{}^<<<<{\alpha_{2n-1}} "f",
\POS "a" \ar@{-}^<<<<{} "g",
\POS "e" \ar@{-}^<<<<{} "g",
\POS "g" \ar@{}^<<{\alpha_{0}} "d",

\endxy$ &$\mathfrak{su}^{(1)}(2n)$ &  & \\ \cline{2-6}
 &$\mathfrak{su}_{-1}^{(1)}(p,q), \newline p+q=2n$ &$\xy

\POS (0,0) *\cir<3pt>{}="a",
\POS (10,0) *\cir<3pt>{} ="b",
\POS (20,0) *\cir<3pt>{} ="c",
\POS (30,0) *\cir<3pt>{} ="d",
\POS (40,0) *\cir<3pt>{} ="e",
\POS (50,0)  ="f"
\POS (20,10) *{\bullet}="g",
\POS "a" \ar@{-}^<<<{\alpha_{1}} "b",
\POS "b" \ar@{.}^<<<{\alpha_{2}} "c",
\POS "c" \ar@{.}^<<<{\alpha_{n}} "d",
\POS "d" \ar@{-}^<<<{\alpha_{2n-2}} "e",
\POS "e" \ar@{}^<<<<{\alpha_{2n-1}} "f",
\POS "a" \ar@{-}^<<<<{} "g",
\POS "e" \ar@{-}^<<<<{} "g",
\POS "g" \ar@{}^<<{\alpha_{0}} "d",

\endxy$ & $\mathfrak{su}(2n)$&$\frac{SU^{(1)}(p+q)}{SU(2n)}$ &$\frac{SU_{-1}^{(1)}(p,q)}{SU(2n)}$ \\ \cline{2-6}
 &$\mathfrak{su}_{1}^{(1)}(p,q),\newline p+q=2n$  &$\xy

\POS (0,0) *\cir<3pt>{}="a",
\POS (10,0) *\cir<3pt>{} ="b",
\POS (20,0) *{\bullet} ="c",
\POS (30,0) *\cir<3pt>{} ="d",
\POS (40,0) *\cir<3pt>{} ="e",
\POS (50,0)  ="f"
\POS (20,10) *{\bullet}="g",
\POS "a" \ar@{-}^<<<{\alpha_{1}} "b",
\POS "b" \ar@{.}^<<<{\alpha_{2}} "c",
\POS "c" \ar@{.}^<<<{\alpha_{n}} "d",
\POS "d" \ar@{-}^<<<{\alpha_{2n-2}} "e",
\POS "e" \ar@{}^<<<<{\alpha_{2n-1}} "f",
\POS "a" \ar@{-}^<<<<{} "g",
\POS "e" \ar@{-}^<<<<{} "g",
\POS "g" \ar@{}^<<{\alpha_{0}} "d",

\endxy$ &$\mathfrak{su}(p)\oplus \mathfrak{su}(q)$ &$\frac{SU^{(1)}(p+q)}{SU(p)\oplus SU(q)}$&$\frac{SU_{1}^{(1)}(p,q)}{SU(p)\oplus SU(q)}$ \\ \cline{2-6}
 &$\mathfrak{sl}_{s}^{(1)}(n,\mathbb{H})$& $\xy
\POS (0,0) * \cir<3pt>{}="a",
\POS (10,0) *\cir<3pt>{} ="b",
\POS (20,0) *\cir<3pt>{} ="c",
\POS (30,0) *\cir<3pt>{} ="d",
\POS (40,0) *\cir<3pt>{} ="e",
\POS (50,0)  ="f"
\POS (20,10) *\cir<3pt>{} ="g",
\POS "a" \ar@{-}^<<<{\alpha_{1}} "b",
\POS "b" \ar@{.}^<<<{\alpha_{2}} "c",
\POS "c" \ar@{.}^<<<{\alpha_{n}} "d",
\POS "d" \ar@{-}^<<<{\alpha_{2n-2}} "e",
\POS "e" \ar@{}^<<<<{\alpha_{2n-1}} "f",
\POS "a" \ar@{-}^<<<<{} "g",
\POS "e" \ar@{-}^<<<<{} "g",
\POS "a" \ar@{<->}@/_{2pc}/ "e",
\POS "b" \ar@{<->}@/_{1pc}/ "d",
\POS "g" \ar@{}^<<{\alpha_{0}} "d",
\endxy$&$\mathfrak{sp}^{(1)}(2n)$ &$\frac{SU^{(1)}(2n)}{SP^{(1)}(2n)}$ &$\frac{SL_{s}^{(1)}(n,\mathbb{H})}{SP^{(1)}(2n)}$ \\ \cline{2-6}
 &$\mathfrak{sl}_{-1}^{(1)}(2n,\mathbb{R}), \newline n\geq 3 $& $\xy
\POS (0,0) * \cir<3pt>{}="a",
\POS (10,0) *\cir<3pt>{} ="b",
\POS (20,0) *\cir<3pt>{} ="c",
\POS (30,0) *\cir<3pt>{} ="d",
\POS (40,0) *\cir<3pt>{} ="e",
\POS (50,0)  ="f"
\POS (20,10) *{\bullet} ="g",
\POS "a" \ar@{-}^<<<{\alpha_{1}} "b",
\POS "b" \ar@{.}^<<<{\alpha_{2}} "c",
\POS "c" \ar@{.}^<<<{\alpha_{n}} "d",
\POS "d" \ar@{-}^<<<{\alpha_{2n-2}} "e",
\POS "e" \ar@{}^<<<<{\alpha_{2n-1}} "f",
\POS "a" \ar@{-}^<<<<{} "g",
\POS "e" \ar@{-}^<<<<{} "g",
\POS "a" \ar@{<->}@/_{2pc}/ "e",
\POS "b" \ar@{<->}@/_{1pc}/ "d",
\POS "g" \ar@{}^<<{\alpha_{0}} "d",

\endxy$&$\mathfrak{su}^{(2)}(2n)$&$\frac{SU^{(1)}(2n)}{SU^{(2)}(2n)}$ &$\frac{SL_{-1}^{(1)}(2n,\mathbb{R})}{SU^{(2)}(2n)}$ \\ \cline{2-6}
 &$\mathfrak{sl}_{1}^{(1)}(2n,\mathbb{R}), \newline n\geq 4$& $\xy
\POS (0,0) * \cir<3pt>{}="a",
\POS (10,0) *\cir<3pt>{} ="b",
\POS (20,0) *{\bullet} ="c",
\POS (30,0) *\cir<3pt>{} ="d",
\POS (40,0) *\cir<3pt>{} ="e",
\POS (50,0)  ="f"
\POS (20,10) *{\bullet} ="g",
\POS "a" \ar@{-}^<<<{\alpha_{1}} "b",
\POS "b" \ar@{.}^<<<{\alpha_{2}} "c",
\POS "c" \ar@{.}^<<<{\alpha_{n}} "d",
\POS "d" \ar@{-}^<<<{\alpha_{2n-2}} "e",
\POS "e" \ar@{}^<<<<{\alpha_{2n-1}} "f",
\POS "a" \ar@{-}^<<<<{} "g",
\POS "e" \ar@{-}^<<<<{} "g",
\POS "a" \ar@{<->}@/_{2pc}/ "e",
\POS "b" \ar@{<->}@/_{1pc}/ "d",
\POS "g" \ar@{}^<<{\alpha_{0}} "d",

\endxy$&$\mathfrak{so}^{(1)}(2n)$ &$\frac{SU^{(1)}(2n)}{SO^{(1)}(2n)}$ &$\frac{SL_{1}^{(1)}(2n,\mathbb{R})}{SO^{(1)}(2n)}$ \\ \cline{2-6}
 &$\mathfrak{sl}_{r^{n}}^{(1)}(n,\mathbb{H})$ & $\xy
\POS (-10,0) ="z"
\POS (0,0) *+\cir<3pt>{} ="a",
\POS (10,0) *+\cir<3pt>{} ="b",
\POS (20,0) *+\cir<3pt>{} ="c",
\POS (-10,-10) *+\cir<3pt>{} ="d",
\POS (30,-10) *+\cir<3pt>{} ="e",
\POS (0,-20) *+\cir<3pt>{} ="f",
\POS (10,-20) *+\cir<3pt>{} ="g",
\POS (20,-20) *++\cir<3pt>{} ="h",

\POS "a" \ar@{.}^<<<{\alpha_{2n-1}} "b",
\POS "b" \ar@{.}^<<<{} "c",
\POS "c" \ar@{-}^<<<{\alpha_{n+1}} "e",
\POS "d" \ar@{-}^<<<{\alpha_{0}} "f",
\POS "f" \ar@{.}^<<<<{\alpha_{1}} "g",
\POS "g" \ar@{.}^<<<<{} "h",
\POS "h" \ar@{-}^<<<<{\alpha_{n-1}} "e",
\POS "a" \ar@{-}^<<<<{} "d",
\POS "e" \ar@{}^<<<{\alpha_{n}} ,
\POS "a" \ar@{<->}"h",
\POS "b" \ar@{<->}"g",
\POS "c" \ar@{<->}"f",
\POS "d" \ar@{<->}"e",
\endxy$&$\mathfrak{su}^{(1)}(n)$ &$\frac{SU^{(1)}(2n)}{SU^{(1)}(n)}$ &$\frac{SL_{r^{n}}^{(1)}(n,\mathbb{H})}{SU^{(1)}(n)}$ \\ \cline{2-6}
 &$\mathfrak{sl}_{rs}^{(1)}(n,\mathbb{H})$ & $\xy
\POS (0,0) *+\cir<3pt>{} ="a",
\POS (10,0) *+\cir<3pt>{} ="b",
\POS (20,0) *+\cir<3pt>{} ="c",
\POS (0,-15) *+\cir<3pt>{} ="d",
\POS (10,-15) *+\cir<3pt>{} ="e",
\POS (20,-15) *+\cir<3pt>{} ="f",

\POS "a" \ar@{.}^<<<{\alpha_{0}} "b",
\POS "c" \ar@{.}^<<{\alpha_{n+1}} "b",
\POS "b" \ar@{<->}^<<<{} "e",
\POS "d" \ar@{.}^<<<{\alpha_{1}} "e",
\POS "f" \ar@{.}^<<{\alpha_{n}} "e",
\POS "c" \ar@{<->}^<<<{} "f",
\POS "a" \ar@{<->}^<<<{} "d",
\POS "a" \ar@{-}@/_{1pc}/ "d"
\POS "c" \ar@{-}@/^{1pc}/ "f"
\endxy$&$\mathfrak{so}^{(2)}(2n)$ &$\frac{SU^{(1)}(2n)}{SO^{(2)}(2n)}$ &$\frac{SL_{rs}^{(1)}(n,\mathbb{H})}{SO^{(2)}(2n)}$ \\ \hline 
\end{tabular}
\end{sidewaystable}

\begin{sidewaystable}[p]
\caption{Affine Kac-Moody symmetric spaces associated with $A_{2n}^{(1)}$}
\centering % centering table
\begin{tabular}{ | l| p{2.5cm} |l | p{2.5cm}| p{2.5cm} |p{2.5cm}|}
\hline
 Dynkin Diagram & Real Forms&Vogan Diagram &Fixed Algebra& Compact affine Kac-Moody Symmetric spaces&Non-compact affine Kac-Moody Symmetric spaces\\
 \hline
 \multirow{8}{*}{$\xy
\POS (0,0) *\cir<3pt>{}="a",
\POS (10,0) *\cir<3pt>{} ="b",
\POS (20,0) *\cir<3pt>{} ="c",
\POS (30,0) *\cir<3pt>{} ="d",
\POS (40,0) *\cir<3pt>{} ="e",
\POS (50,0)  ="f"
\POS (20,10) *\cir<3pt>{} ="g",
\POS "a" \ar@{-}^<<<{\alpha_{1}} "b",
\POS "b" \ar@{.}^<<<{\alpha_{2}} "c",
\POS "c" \ar@{.}^<<<{\alpha_{n}} "d",
\POS "d" \ar@{-}^<<<{\alpha_{2n-1}} "e",
\POS "e" \ar@{}^<<<<{\alpha_{2n}} "f",
\POS "a" \ar@{-}^<<<<{} "g",
\POS "e" \ar@{-}^<<<<{} "g",
\POS "g" \ar@{}^<<{\alpha_{0}} "d",
\endxy$ } &$\mathfrak{su}^{(1)}(2n+1)$ & $\xy
\POS (0,0) *\cir<3pt>{}="a",
\POS (10,0) *\cir<3pt>{} ="b",
\POS (20,0) *\cir<3pt>{} ="c",
\POS (30,0) *\cir<3pt>{} ="d",
\POS (40,0) *\cir<3pt>{} ="e",
\POS (50,0)  ="f"
\POS (20,10) *\cir<3pt>{} ="g",
\POS "a" \ar@{-}^<<<{\alpha_{1}} "b",
\POS "b" \ar@{.}^<<<{\alpha_{2}} "c",
\POS "c" \ar@{.}^<<<{\alpha_{n}} "d",
\POS "d" \ar@{-}^<<<{\alpha_{2n-1}} "e",
\POS "e" \ar@{}^<<<<{\alpha_{2n}} "f",
\POS "a" \ar@{-}^<<<<{} "g",
\POS "e" \ar@{-}^<<<<{} "g",
\POS "g" \ar@{}^<<{\alpha_{0}} "d",
\endxy$ &$\mathfrak{su}^{(1)}(2n+1)$& &\\ \cline{2-6}
 &$\mathfrak{su}_{-1}^{(1)}(p,q),\newline p+q=2n+1$ & $\xy

\POS (0,0) *\cir<3pt>{}="a",
\POS (10,0) *\cir<3pt>{} ="b",
\POS (20,0) *\cir<3pt>{} ="c",
\POS (30,0) *\cir<3pt>{} ="d",
\POS (40,0) *\cir<3pt>{} ="e",
\POS (50,0)  ="f"
\POS (20,10) *{\bullet}="g",
\POS "a" \ar@{-}^<<<{\alpha_{1}} "b",
\POS "b" \ar@{.}^<<<{\alpha_{2}} "c",
\POS "c" \ar@{.}^<<<{\alpha_{n}} "d",
\POS "d" \ar@{-}^<<<{\alpha_{2n-1}} "e",
\POS "e" \ar@{}^<<<<{\alpha_{2n}} "f",
\POS "a" \ar@{-}^<<<<{} "g",
\POS "e" \ar@{-}^<<<<{} "g",
\POS "g" \ar@{}^<<{\alpha_{0}} "d",

\endxy$&$\mathfrak{su}(2n+1)$ &$\frac{SU^{(1)}(p+q)}{SU(2n+1)}$ &$\frac{SU_{-1}^{(1)}(p,q)}{SU(2n+1)}$ \\ \cline{2-6}
 &$\mathfrak{su}_{1}^{(1)}(p,q),\newline p+q=2n+1$ & $\xy

\POS (0,0) *\cir<3pt>{}="a",
\POS (10,0) *\cir<3pt>{} ="b",
\POS (20,0) *{\bullet}="c",
\POS (30,0) *\cir<3pt>{} ="d",
\POS (40,0) *\cir<3pt>{} ="e",
\POS (50,0)  ="f"
\POS (20,10) *{\bullet}="g",
\POS "a" \ar@{-}^<<<{\alpha_{1}} "b",
\POS "b" \ar@{.}^<<<{\alpha_{2}} "c",
\POS "c" \ar@{.}^<<<{\alpha_{n}} "d",
\POS "d" \ar@{-}^<<<{\alpha_{2n-1}} "e",
\POS "e" \ar@{}^<<<<{\alpha_{2n}} "f",
\POS "a" \ar@{-}^<<<<{} "g",
\POS "e" \ar@{-}^<<<<{} "g",
\POS "g" \ar@{}^<<{\alpha_{0}} "d",

\endxy$&$\mathfrak{su}(p)\oplus \mathfrak{su}(q)$ &$\frac{SU^{(1)}(p+q)}{SU(p)\oplus SU(q)}$ &$\frac{SU_{1}^{(1)}(p,q)}{SU(p)\oplus SU(q)}$ \\ \cline{2-6}
 &$\mathfrak{sl}_{-1}^{(1)}(2n+1,\mathbb{R})$ &$\xy
\POS (0,0) * \cir<3pt>{}="a",
\POS (10,0) *\cir<3pt>{} ="b",
\POS (20,0) *\cir<3pt>{} ="c",
\POS (30,0) *\cir<3pt>{} ="d",
\POS (40,0) *\cir<3pt>{} ="e",
\POS (50,0)  ="f"
\POS (20,10) *\cir<3pt>{} ="g",
\POS "a" \ar@{-}^<<<{\alpha_{1}} "b",
\POS "b" \ar@{.}^<<<{\alpha_{2}} "c",
\POS "c" \ar@{.}^<<<{\alpha_{n}} "d",
\POS "d" \ar@{-}^<<<{\alpha_{2n-1}} "e",
\POS "e" \ar@{}^<<<<{\alpha_{2n}} "f",
\POS "a" \ar@{-}^<<<<{} "g",
\POS "e" \ar@{-}^<<<<{} "g",
\POS "a" \ar@{<->}@/_{2pc}/ "e",
\POS "b" \ar@{<->}@/_{1pc}/ "d",
\POS "g" \ar@{}^<<{\alpha_{0}} "d",

\endxy$ &$\mathfrak{su}^{(2)}(2n+1)$ &$\frac{SU^{(1)}(2n+1)}{SU^{(2)}(2n+1)}$ &$\frac{SL_{-1}^{(1)}(2n+1,\mathbb{R})}{SU^{(2)}(2n+1)}$ \\ \cline{2-6}
 &$\mathfrak{sl}_{1}^{(1)}(2n+,\mathbb{R}),\newline n\geq 3 $ &$\xy
\POS (0,0) * \cir<3pt>{}="a",
\POS (10,0) *\cir<3pt>{} ="b",
\POS (20,0) *\cir<3pt>{} ="c",
\POS (30,0) *\cir<3pt>{} ="d",
\POS (40,0) *\cir<3pt>{} ="e",
\POS (50,0)  ="f"
\POS (20,10) *{\bullet} ="g",
\POS "a" \ar@{-}^<<<{\alpha_{1}} "b",
\POS "b" \ar@{.}^<<<{\alpha_{2}} "c",
\POS "c" \ar@{.}^<<<{\alpha_{n}} "d",
\POS "d" \ar@{-}^<<<{\alpha_{2n-2}} "e",
\POS "e" \ar@{}^<<<<{\alpha_{2n-1}} "f",
\POS "a" \ar@{-}^<<<<{} "g",
\POS "e" \ar@{-}^<<<<{} "g",
\POS "a" \ar@{<->}@/_{2pc}/ "e",
\POS "b" \ar@{<->}@/_{1pc}/ "d",
\POS "g" \ar@{}^<<{\alpha_{0}} "d",

\endxy$  &$\mathfrak{so}^{(1)}(2n)$ &$\frac{SU^{(1)}(2n+1)}{SO^{(1)}(2n)}$ &$\frac{SL_{1}^{(1)}(2n+1,\mathbb{R})}{SO^{(1)}(2n)}$ \\ \hline
 \end{tabular}
\end{sidewaystable}
\begin{sidewaystable}[p]
\caption{Affine Kac-Moody symmetric spaces associated with $B_{n}^{(1)}$}
\centering % centering table
\begin{tabular}{ | l| p{2.9cm} |l | p{3cm}| p{2.7cm} |p{2.7cm}|}
\hline
 Dynkin Diagram & Real Forms&Vogan Diagram &Fixed Algebra& Compact affine Kac-Moody Symmetric spaces&Non-compact affine Kac-Moody Symmetric spaces\\
 \hline
 \multirow{9}{*}{$\xy
\POS (0,0) *\cir<3pt>{} ="a",
\POS (5,0) *\cir<3pt>{} ="b",
\POS (10,0) *\cir<3pt>{} ="c",
\POS (15,0) *\cir<3pt>{} ="e",
\POS (20,0) *\cir<3pt>{} ="g",
\POS (-8,8) *\cir<3pt>{} ="d",
\POS (-8,-8) *\cir<3pt>{} ="f",
\POS "a" \ar@{-}^<<<{\alpha_{2}} "b",
\POS "b" \ar@{.}^<<<{\alpha_{3}} "c",
\POS "c" \ar@{.}^<<{\alpha_{p}} "e",
\POS "e" \ar@{=>}^<<{\alpha_{n-1}} "g",
\POS "d" \ar@{-}^<<<{\alpha_{0}} "a",
\POS "f" \ar@{-}^<<{\alpha_{1}} "a",
\POS "g" \ar@{}^<<{\alpha_{n}},
\endxy$} & $\mathfrak{so}^{(1)}(2n+1)$ & $\xy
\POS (0,0) *\cir<3pt>{} ="a",
\POS (10,0) *\cir<3pt>{} ="b",
\POS (20,0) *\cir<3pt>{} ="c",
\POS (30,0) *\cir<3pt>{} ="e",
\POS (40,0) *\cir<3pt>{} ="g",
\POS (-5,5) *\cir<3pt>{} ="d",
\POS (-5,-5) *\cir<3pt>{} ="f",
\POS "a" \ar@{-}^<<<{\alpha_{2}} "b",
\POS "b" \ar@{.}^<<<{\alpha_{3}} "c",
\POS "c" \ar@{.}^<<{\alpha_{p}} "e",
\POS "e" \ar@{=>}^<<{\alpha_{n-1}} "g",
\POS "d" \ar@{-}^<<<{\alpha_{0}} "a",
\POS "f" \ar@{-}^<<{\alpha_{1}} "a",
\POS "g" \ar@{}^<<{\alpha_{n}},
\endxy$ & $\mathfrak{so}^{(1)}(2n+1)$ &  &\\ \cline{2-6}
 &$\mathfrak{so}_{-1}^{(1)}(2,2n-1)$ &$\xy
\POS (0,0) *\cir<3pt>{} ="a",
\POS (10,0) *\cir<3pt>{} ="b",
\POS (20,0) *\cir<3pt>{} ="c",
\POS (30,0) *\cir<3pt>{} ="e",
\POS (40,0) *\cir<3pt>{} ="g",
\POS (-5,5) *\cir<3pt>{} ="d",
\POS (-5,-5) *{\bullet}="f",
\POS "a" \ar@{-}^<<<{\alpha_{2}} "b",
\POS "b" \ar@{.}^<<<{\alpha_{3}} "c",
\POS "c" \ar@{.}^<<{\alpha_{p}} "e",
\POS "e" \ar@{=>}^<<{\alpha_{n-1}} "g",
\POS "d" \ar@{-}^<<<{\alpha_{0}} "a",
\POS "f" \ar@{-}^<<{\alpha_{1}} "a",
\POS "g" \ar@{}^<<{\alpha_{n}},
\endxy$  &$\mathfrak{so}(2n+1)$ &$\frac{SO^{(1)}(2n+1)}{SO(2n+1)}$ &$\frac{SO_{-1}^{(1)}(2,2n-1)}{SO(2n+1)}$ \\ \cline{2-6}
 &$\mathfrak{so}^{(1)}(4,2n-3)$ &$\xy
\POS (0,0) *{\bullet} ="a",
\POS (10,0) *\cir<3pt>{} ="b",
\POS (20,0) *\cir<3pt>{} ="c",
\POS (30,0) *\cir<3pt>{} ="e",
\POS (40,0) *\cir<3pt>{} ="g",
\POS (-5,5) *\cir<3pt>{} ="d",
\POS (-5,-5) *\cir<3pt>{} ="f",
\POS "a" \ar@{-}^<<<{\alpha_{2}} "b",
\POS "b" \ar@{.}^<<<{\alpha_{3}} "c",
\POS "c" \ar@{.}^<<{\alpha_{p}} "e",
\POS "e" \ar@{=>}^<<{\alpha_{n-1}} "g",
\POS "d" \ar@{-}^<<<{\alpha_{0}} "a",
\POS "f" \ar@{-}^<<{\alpha_{1}} "a",
\POS "g" \ar@{}^<<{\alpha_{n}},
\endxy$ &$\mathfrak{so}(4)\oplus \mathfrak{so}(2n-3)$&$\frac{SO^{(1)}(2n+1)}{SO(4)\oplus SO(2n-3)}$ &$\frac{SO^{(1)}(4,2n-3)}{SO(4)\oplus SO(2n-3)}$  \\ \cline{2-6}
 &$\mathfrak{so}^{(1)}(6,2n-5)$ &$\xy
\POS (0,0) *\cir<3pt>{} ="a",
\POS (10,0) * {\bullet}="b",
\POS (20,0) *\cir<3pt>{} ="c",
\POS (30,0) *\cir<3pt>{} ="e",
\POS (40,0) *\cir<3pt>{} ="g",
\POS (-5,5) *\cir<3pt>{} ="d",
\POS (-5,-5) *\cir<3pt>{} ="f",
\POS "a" \ar@{-}^<<<{\alpha_{2}} "b",
\POS "b" \ar@{.}^<<<{\alpha_{3}} "c",
\POS "c" \ar@{.}^<<{\alpha_{p}} "e",
\POS "e" \ar@{=>}^<<{\alpha_{n-1}} "g",
\POS "d" \ar@{-}^<<<{\alpha_{0}} "a",
\POS "f" \ar@{-}^<<{\alpha_{1}} "a",
\POS "g" \ar@{}^<<{\alpha_{n}},
\endxy$ &$\mathfrak{su}^{(1)}(4)\oplus \mathfrak{so}(2n-5)$&$\frac{SO^{(1)}(2n+1)}{SU^{(1)}(4)\oplus SO(2n-5)}$ &$\frac{SO^{(1)}(6,2n-5)}{SU^{(1)}(4)\oplus SO(2n-5)}$  \\ \cline{2-6}

 &$\mathfrak{so}^{(1)}(2p,2q+1)\newline p+q=n$ &$\xy
\POS (0,0) *\cir<3pt>{} ="a",
\POS (10,0) *\cir<3pt>{}="b",
\POS (20,0) *{\bullet}="c",
\POS (30,0) *\cir<3pt>{}="e",
\POS (40,0) *\cir<3pt>{} ="g",
\POS (-5,5) *\cir<3pt>{} ="d",
\POS (-5,-5) *\cir<3pt>{} ="f",
\POS "a" \ar@{-}^<<<{\alpha_{2}} "b",
\POS "b" \ar@{.}^<<<{\alpha_{3}} "c",
\POS "c" \ar@{.}^<<{\alpha_{p}} "e",
\POS "e" \ar@{=>}^<<{\alpha_{n-1}} "g",
\POS "d" \ar@{-}^<<<{\alpha_{0}} "a",
\POS "f" \ar@{-}^<<{\alpha_{1}} "a",
\POS "g" \ar@{}^<<{\alpha_{n}},
\endxy$ &$\mathfrak{so}^{(1)}(2p)\oplus \mathfrak{so}(2q+1)$&$\frac{SO^{(1)}(2n+1)}{SO^{(1)}(2p)\oplus SO(2q+1)}$ &$\frac{SO^{(1)}(2p,2q+1)}{SO^{(1)}(2p)\oplus SO(2q+1)}$  \\ \cline{2-6}
 &$\mathfrak{so}^{(1)}(2n,1)$ &$\xy
\POS (0,0) *\cir<3pt>{} ="a",
\POS (10,0) *\cir<3pt>{}="b",
\POS (20,0) *\cir<3pt>{} ="c",
\POS (30,0) *\cir<3pt>{}="e",
\POS (40,0) *{\bullet}  ="g",
\POS (-5,5) *\cir<3pt>{} ="d",
\POS (-5,-5) *\cir<3pt>{} ="f",
\POS "a" \ar@{-}^<<<{\alpha_{2}} "b",
\POS "b" \ar@{.}^<<<{\alpha_{3}} "c",
\POS "c" \ar@{.}^<<{\alpha_{p}} "e",
\POS "e" \ar@{=>}^<<{\alpha_{n-1}} "g",
\POS "d" \ar@{-}^<<<{\alpha_{0}} "a",
\POS "f" \ar@{-}^<<{\alpha_{1}} "a",
\POS "g" \ar@{}^<<{\alpha_{n}},
\endxy$ &$\mathfrak{so}^{(1)}(2n)$&$\frac{SO^{(1)}(2n+1)}{SO^{(1)}(2n)}$ &$\frac{SO^{(1)}(2n,1)}{SO^{(1)}(2n)}$  \\ \cline{2-6} 
 &$\mathfrak{so}_{1}^{(1)}(2,2n-1)$ &$\xy
\POS (0,0) *\cir<3pt>{} ="a",
\POS (10,0) *\cir<3pt>{}="b",
\POS (20,0) *\cir<3pt>{} ="c",
\POS (30,0) *\cir<3pt>{}="e",
\POS (40,0) *\cir<3pt>{}  ="g",
\POS (-5,5) *{\bullet}="d",
\POS (-5,-5) *{\bullet} ="f",
\POS "a" \ar@{-}^<<<{\alpha_{2}} "b",
\POS "b" \ar@{.}^<<<{\alpha_{3}} "c",
\POS "c" \ar@{.}^<<{\alpha_{p}} "e",
\POS "e" \ar@{=>}^<<{\alpha_{n-1}} "g",
\POS "d" \ar@{-}^<<<{\alpha_{0}} "a",
\POS "f" \ar@{-}^<<{\alpha_{1}} "a",
\POS "g" \ar@{}^<<{\alpha_{n}},
\endxy$ &$\mathfrak{so}(2n-1)$&$\frac{SO^{(1)}(2n+1)}{SO(2n-1)}$ &$\frac{SO_{1}^{(1)}(2,2n-1)}{SO(2n-1)}$  \\ \cline{2-6} 
 &$\mathfrak{so}^{(1)}(1,2n)$ &$\xy
\POS (0,0) *\cir<3pt>{} ="a",
\POS (10,0) *\cir<3pt>{} ="b",
\POS (20,0) *\cir<3pt>{} ="c",
\POS (30,0) *\cir<3pt>{} ="e",
\POS (40,0) *\cir<3pt>{} ="g",
\POS (-5,5) *\cir<3pt>{} ="d",
\POS (-5,-5) *\cir<3pt>{} ="f",
\POS "a" \ar@{-}^<<<{\alpha_{2}} "b",
\POS "b" \ar@{.}^<<<{\alpha_{3}} "c",
\POS "c" \ar@{.}^<<{\alpha_{p}} "e",
\POS "e" \ar@{=>}^<<{\alpha_{n-1}} "g",
\POS "d" \ar@{-}^<<<{\alpha_{0}} "a",
\POS "f" \ar@{-}^<<{\alpha_{1}} "a",
\POS "g" \ar@{}^<<{\alpha_{n}},
\POS "d" \ar@{<->}@/_{1pc}/ "f",
\endxy$ &$\mathfrak{so}^{(2)}(2n)$ &$\frac{SO^{(1)}(2n+1)}{SO^{(2)}(2n)}$ &$\frac{SO^{(1)}(1,2n)}{SO^{(2)}(2n)}$ \\ \cline{2-6}
&$\mathfrak{so}^{(1)}(5,2n-4) $&$\xy
\POS (0,0) *{\bullet} ="a",
\POS (10,0) *\cir<3pt>{}="b",
\POS (20,0) *\cir<3pt>{} ="c",
\POS (30,0) *\cir<3pt>{} ="e",
\POS (40,0) *\cir<3pt>{} ="g",
\POS (-5,5) *\cir<3pt>{} ="d",
\POS (-5,-5) *\cir<3pt>{} ="f",
\POS "a" \ar@{-}^<<<{\alpha_{2}} "b",
\POS "b" \ar@{.}^<<<{\alpha_{3}} "c",
\POS "c" \ar@{.}^<<{\alpha_{p}} "e",
\POS "e" \ar@{=>}^<<{\alpha_{n-1}} "g",
\POS "d" \ar@{-}^<<<{\alpha_{0}} "a",
\POS "f" \ar@{-}^<<{\alpha_{1}} "a",
\POS "g" \ar@{}^<<{\alpha_{n}},
\POS "d" \ar@{<->}@/_{1pc}/ "f",
\endxy$ &$\mathfrak{su}(3)\oplus\mathfrak{so}(2n-3)$ &$\frac{SO^{(1)}(2n+1)}{SU(3)\oplus SO(2n-3)}$ &$\frac{SO^{(1)}(3,2n-2)}{SU(3)\oplus SO(2n-3)}$\\ \cline{2-6}
&$\mathfrak{so}^{(1)}(2p+1,2q)\newline p+q=n $ &$\xy
\POS (0,0) *\cir<3pt>{} ="a",
\POS (10,0) *\cir<3pt>{}="b",
\POS (20,0) *{\bullet}="c",
\POS (30,0) *\cir<3pt>{} ="e",
\POS (40,0) *\cir<3pt>{} ="g",
\POS (-5,5) *\cir<3pt>{} ="d",
\POS (-5,-5) *\cir<3pt>{} ="f",
\POS "a" \ar@{-}^<<<{\alpha_{2}} "b",
\POS "b" \ar@{.}^<<<{\alpha_{3}} "c",
\POS "c" \ar@{.}^<<{\alpha_{p}} "e",
\POS "e" \ar@{=>}^<<{\alpha_{n-1}} "g",
\POS "d" \ar@{-}^<<<{\alpha_{0}} "a",
\POS "f" \ar@{-}^<<{\alpha_{1}} "a",
\POS "d" \ar@{<->}@/_{1pc}/ "f",
\POS "g" \ar@{}^<<{\alpha_{n}},
\endxy$ &$\mathfrak{so}^{(2)}(2p)\oplus\mathfrak{so}(2q+1)$ &$\frac{SO^{(1)}(2n+1)}{SO^{(2)}(2p)\oplus SO(2q+1)}$  &$\frac{SO^{(1)}(2p+1,2q)}{SO^{(2)}(2p)\oplus SO(2q+1)}$ \\ \cline{2-6}
&$\mathfrak{so}^{(1)}(2n-3,4)$ &$\xy
\POS (0,0) *\cir<3pt>{} ="a",
\POS (10,0) *\cir<3pt>{}="b",
\POS (20,0) *\cir<3pt>{} ="c",
\POS (30,0) *\cir<3pt>{}="e",
\POS (40,0) *{\bullet}  ="g",
\POS (-5,5) *\cir<3pt>{} ="d",
\POS (-5,-5) *\cir<3pt>{} ="f",
\POS "a" \ar@{-}^<<<{\alpha_{2}} "b",
\POS "b" \ar@{.}^<<<{\alpha_{3}} "c",
\POS "c" \ar@{.}^<<{\alpha_{p}} "e",
\POS "e" \ar@{=>}^<<{\alpha_{n-1}} "g",
\POS "d" \ar@{-}^<<<{\alpha_{0}} "a",
\POS "f" \ar@{-}^<<{\alpha_{1}} "a",
\POS "d" \ar@{<->}@/_{1pc}/ "f",
\POS "g" \ar@{}^<<{\alpha_{n}},
\endxy$&$\mathfrak{so}^{(2)}(2n)$ &$\frac{SO^{(1)}(2n+1)}{SO^{(2)}(2n)}$ &$\frac{SO^{(1)}(2n-3,4)}{SO^{(2)}(2n)}$ \\ \cline{2-6}
\hline
  
\end{tabular}
\end{sidewaystable}

\begin{sidewaystable}[p]
\caption{Affine Kac-Moody symmetric spaces associated with $C_{2n-1}^{(1)}$}
\centering % centering table
\begin{tabular}{ | l| p{3cm} |l | p{3cm}| p{2.5cm} |p{2.5cm}|}
\hline
 Dynkin Diagram & Real Forms&Vogan Diagram &Fixed Algebra& Compact affine Kac-Moody Symmetric spaces&Non-compact affine Kac-Moody Symmetric spaces\\
 \hline
 \multirow{7}{*}{$\xy
\POS (0,0) *\cir<3pt>{}="a",
\POS (7,0) *\cir<3pt>{} ="b",
\POS (14,0) *\cir<3pt>{} ="c",
\POS (21,0) *\cir<3pt>{} ="d",
\POS (28,0) *\cir<3pt>{} ="e",
\POS (35,0) *\cir<3pt>{} ="f"
\POS "a" \ar@{=>}^<<<{\alpha_{0}} "b",
\POS "b" \ar@{.}^<<<{\alpha_{1}} "c",
\POS "c" \ar@{-}^<<<{\alpha_{n-1}} "d",
\POS "d" \ar@{.}^<<<{\alpha_{n}} "e",
\POS "e" \ar@{<=}^<<<<{\alpha_{2n-2}} "f",
\POS "f" \ar@{}^<<{\alpha_{2n-1}},
\endxy$} & $\mathfrak{sp}^{(1)}(2n-1)$ &$\xy
\POS (0,0) *\cir<3pt>{}="a",
\POS (10,0) *\cir<3pt>{} ="b",
\POS (20,0) *\cir<3pt>{} ="c",
\POS (30,0) *\cir<3pt>{} ="d",
\POS (40,0) *\cir<3pt>{} ="e",
\POS (50,0) *\cir<3pt>{} ="f"
\POS "a" \ar@{=>}^<<<{\alpha_{0}} "b",
\POS "b" \ar@{.}^<<<{\alpha_{1}} "c",
\POS "c" \ar@{-}^<<<{\alpha_{p-1}} "d",
\POS "d" \ar@{.}^<<<{\alpha_{p}} "e",
\POS "e" \ar@{<=}^<<<<{\alpha_{2n-2}} "f",
\POS "f" \ar@{}^<<{\alpha_{2n-1}},
\endxy$ & $\mathfrak{sp}^{(1)}(2n-1)$ & &\\ \cline{2-6}
 &$\mathfrak{sp}^{(1)}(p,q)\newline p+q=2n-1$ &$\xy
\POS (0,0) *\cir<3pt>{}="a",
\POS (10,0) *\cir<3pt>{} ="b",
\POS (20,0) *\cir<3pt>{} ="c",
\POS (30,0) *{\bullet} ="d",
\POS (40,0) *\cir<3pt>{} ="e",
\POS (50,0) *\cir<3pt>{} ="f"
\POS "a" \ar@{=>}^<<<{\alpha_{0}} "b",
\POS "b" \ar@{.}^<<<{\alpha_{1}} "c",
\POS "c" \ar@{-}^<<<{\alpha_{p-1}} "d",
\POS "d" \ar@{.}^<<<{\alpha_{p}} "e",
\POS "e" \ar@{<=}^<<<<{\alpha_{2n-2}} "f",
\POS "f" \ar@{}^<<{\alpha_{2n-1}},
\endxy$ &$\mathfrak{sp}^{(1)}(p)\oplus\mathfrak{sp}(q) $  &$\frac{SP^{(1)}(p+q)}{SP^{(1)}(p)\oplus SP(q)}$ & $\frac{SP^{(1)}(p,q)}{SP^{(1)}(p)\oplus SP(q)}$\\ \cline{2-6}
 &$\mathfrak{sp}_{-1}^{(1)}(2n-1,\mathbb{R})$ &$\xy
\POS (0,0) *\cir<3pt>{}="a",
\POS (10,0) *\cir<3pt>{} ="b",
\POS (20,0) *\cir<3pt>{} ="c",
\POS (30,0) *\cir<3pt>{}="d",
\POS (40,0) *\cir<3pt>{}="e",
\POS (50,0) *{\bullet} ="f"
\POS "a" \ar@{=>}^<<<{\alpha_{0}} "b",
\POS "b" \ar@{.}^<<<{\alpha_{1}} "c",
\POS "c" \ar@{-}^<<<{\alpha_{p-1}} "d",
\POS "d" \ar@{.}^<<<{\alpha_{p}} "e",
\POS "e" \ar@{<=}^<<<<{\alpha_{2n-2}} "f",
\POS "f" \ar@{}^<<{\alpha_{2n-1}},
\endxy$ &$\mathfrak{sp}(2n-1)$ &$\frac{SP^{(1)}(2n-1)}{SP(2n-1)}$ & $\frac{SP_{-1}^{(1)}(2n-1,\mathbb{R})}{SP(2n-1)}$\\ \cline{2-6}
 &$\mathfrak{sp}_{1}^{(1)}(2n-1,\mathbb{R})$ &$\xy
\POS (0,0) *{\bullet}="a",
\POS (10,0) *\cir<3pt>{} ="b",
\POS (20,0) *\cir<3pt>{} ="c",
\POS (30,0) *\cir<3pt>{}="d",
\POS (40,0) *\cir<3pt>{}="e",
\POS (50,0) *{\bullet} ="f"
\POS "a" \ar@{=>}^<<<{\alpha_{0}} "b",
\POS "b" \ar@{.}^<<<{\alpha_{1}} "c",
\POS "c" \ar@{-}^<<<{\alpha_{p-1}} "d",
\POS "d" \ar@{.}^<<<{\alpha_{p}} "e",
\POS "e" \ar@{<=}^<<<<{\alpha_{2n-2}} "f",
\POS "f" \ar@{}^<<{\alpha_{2n-1}},
\endxy$ &$\mathfrak{su}(2n-1)$ &$\frac{SP^{(1)}(2n-1)}{SU(2n-1)}$ & $\frac{SP_{1}^{(1)}(2n-1,\mathbb{R})}{SU(2n-1)}$\\ \cline{2-6}
 &$\mathfrak{sp}^{(1)}(2n-1,\mathbb{R})$ &$\xy
\POS (0,0) *\cir<3pt>{}="a",
\POS (10,0) *\cir<3pt>{} ="b",
\POS (20,0) *\cir<3pt>{} ="c",
\POS (30,0) *\cir<3pt>{} ="d",
\POS (40,0) *\cir<3pt>{} ="e",
\POS (50,0) *\cir<3pt>{} ="f"
\POS "a" \ar@{=>}^<<<{\alpha_{0}} "b",
\POS "b" \ar@{.}^<<<{\alpha_{1}} "c",
\POS "c" \ar@{-}^<<<{\alpha_{p-1}} "d",
\POS "d" \ar@{.}^<<<{\alpha_{p}} "e",
\POS "e" \ar@{<=}^<<<<{\alpha_{2n-2}} "f",
\POS "f" \ar@{}^<<{\alpha_{2n-1}},
\POS "a" \ar@{<->}@/_{1.5pc}/ "f",
\POS "b" \ar@{<->}@/_{1pc}/ "e",
\POS "c" \ar@{<->}@/_{.5pc}/ "d",
\endxy$  &$\mathfrak{su}^{(2)}(2n-1)$ &$\frac{SP^{(1)}(2n-1)}{SU^{(2)}(2n-1)}$ &$\frac{SP^{(1)}(2n-1,\mathbb{R})}{SU^{(2)}(2n-1)}$ \\ 
  \hline 
\end{tabular}
\end{sidewaystable} 

\begin{sidewaystable}[p]
\caption{Affine Kac-Moody symmetric spaces associated with $C_{2n}^{(1)}$}
\centering % centering table
\begin{tabular}{ | l| p{2.6cm} |l | p{3cm}| p{2.5cm} |p{2cm}|}
\hline
 Dynkin Diagram & Real Forms&Vogan Diagram &Fixed Algebra& Compact affine Kac-Moody Symmetric spaces&Non-compact affine Kac-Moody Symmetric spaces\\
 \hline
 \multirow{7}{*}{$\xy
\POS (0,0) *\cir<3pt>{}="a",
\POS (6,0) *\cir<3pt>{} ="b",
\POS (12,0) *\cir<3pt>{} ="c",
\POS (18,0) *\cir<3pt>{} ="d",
\POS (24,0) *\cir<3pt>{} ="e",
\POS (30,0) *\cir<3pt>{} ="f"
\POS (36,0) *\cir<3pt>{} ="g"
\POS "a" \ar@{=>}^<<<{\alpha_{0}} "b",
\POS "b" \ar@{.}^<<<{\alpha_{1}} "c",
\POS "c" \ar@{-}_<<<{\alpha_{n-1}} "d",
\POS "d" \ar@{-}^<<<{\alpha_{n}} "e",
\POS "e" \ar@{.}_<<<<{\alpha_{n+1}} "f",
\POS "f" \ar@{<=}^<<<<{\alpha_{2n-1}} "g",
\POS "g" \ar@{}^<<{\alpha_{2n}},
\endxy$} & $\mathfrak{sp}^{(1)}(2n)$ &$\xy
\POS (0,0) *\cir<3pt>{}="a",
\POS (10,0) *\cir<3pt>{} ="b",
\POS (20,0) *\cir<3pt>{} ="c",
\POS (30,0) *\cir<3pt>{} ="d",
\POS (40,0) *\cir<3pt>{} ="e",
\POS (50,0) *\cir<3pt>{} ="f"
\POS (60,0) *\cir<3pt>{} ="g"
\POS "a" \ar@{=>}^<<<{\alpha_{0}} "b",
\POS "b" \ar@{.}^<<<{\alpha_{1}} "c",
\POS "c" \ar@{-}^<<<{\alpha_{n-1}} "d",
\POS "d" \ar@{-}^<<<{\alpha_{n}} "e",
\POS "e" \ar@{.}^<<<<{\alpha_{n+1}} "f",
\POS "f" \ar@{<=}^<<<<{\alpha_{2n-1}} "g",
\POS "g" \ar@{}^<<{\alpha_{2n}},
\endxy$ & $\mathfrak{sp}^{(1)}(2n)$ & &\\ \cline{2-6}
 &$\mathfrak{sp}^{(1)}(p,q)\newline p+q=2n,\; p>1$ &$\xy
\POS (0,0) *\cir<3pt>{}="a",
\POS (10,0) *\cir<3pt>{} ="b",
\POS (20,0) *\cir<3pt>{} ="c",
\POS (30,0) *{\bullet} ="d",
\POS (40,0) *\cir<3pt>{} ="e",
\POS (50,0) *\cir<3pt>{} ="f"
\POS (60,0) *\cir<3pt>{} ="g"
\POS "a" \ar@{=>}^<<<{\alpha_{0}} "b",
\POS "b" \ar@{.}^<<<{\alpha_{1}} "c",
\POS "c" \ar@{-}^<<<{\alpha_{p-1}} "d",
\POS "d" \ar@{-}^<<<{\alpha_{p}} "e",
\POS "e" \ar@{.}^<<<<{\alpha_{p+1}} "f",
\POS "f" \ar@{<=}^<<<<{\alpha_{2n-1}} "g",
\POS "g" \ar@{}^<<{\alpha_{2n}},
\endxy$&$\mathfrak{sp}^{(1)}(p)\oplus\mathfrak{sp}(q)$ &$\frac{SP^{(1)}(p+q)}{SP^{(1)}(p)\oplus SP(q)}$&$\frac{SP^{(1)}(p,q)}{SP^{(1)}(p)\oplus SP(q)}$ \\ \cline{2-6}

&$\mathfrak{sp}_{-1}^{(1)}(2n,\mathbb{R})$ &$\xy
\POS (0,0) *\cir<3pt>{}="a",
\POS (10,0) *\cir<3pt>{} ="b",
\POS (20,0) *\cir<3pt>{} ="c",
\POS (30,0) *\cir<3pt>{} ="d",
\POS (40,0) *\cir<3pt>{} ="e",
\POS (50,0) *\cir<3pt>{} ="f"
\POS (60,0) *{\bullet} ="g"
\POS "a" \ar@{=>}^<<<{\alpha_{0}} "b",
\POS "b" \ar@{.}^<<<{\alpha_{1}} "c",
\POS "c" \ar@{-}^<<<{\alpha_{n-1}} "d",
\POS "d" \ar@{-}^<<<{\alpha_{n}} "e",
\POS "e" \ar@{.}^<<<<{\alpha_{n+1}} "f",
\POS "f" \ar@{<=}^<<<<{\alpha_{2n-1}} "g",
\POS "g" \ar@{}^<<{\alpha_{2n}},
\endxy$ &$\mathfrak{sp}(2n)$  &$\frac{SP^{(1)}(2n)}{SP(2n)}$  &$\frac{SP_{-1}^{(1)}(2n,\mathbb{R})}{SP(2n)}$  \\ \cline{2-6}
&$\mathfrak{sp}_{1}^{(1)}(2n,\mathbb{R})$ &$\xy
\POS (0,0) *{\bullet}="a",
\POS (10,0) *\cir<3pt>{} ="b",
\POS (20,0) *\cir<3pt>{} ="c",
\POS (30,0) *\cir<3pt>{} ="d",
\POS (40,0) *\cir<3pt>{} ="e",
\POS (50,0) *\cir<3pt>{} ="f"
\POS (60,0) *{\bullet} ="g"
\POS "a" \ar@{=>}^<<<{\alpha_{0}} "b",
\POS "b" \ar@{.}^<<<{\alpha_{1}} "c",
\POS "c" \ar@{-}^<<<{\alpha_{n-1}} "d",
\POS "d" \ar@{-}^<<<{\alpha_{n}} "e",
\POS "e" \ar@{.}^<<<<{\alpha_{n+1}} "f",
\POS "f" \ar@{<=}^<<<<{\alpha_{2n-1}} "g",
\POS "g" \ar@{}^<<{\alpha_{2n}},
\endxy$ &$\mathfrak{su}(2n)$   &$\frac{SP^{(1)}(2n)}{SU(2n)}$  &$\frac{SP_{1}^{(1)}(2n,\mathbb{R})}{SU(2n)}$  \\ \cline{2-6}
 &$\mathfrak{sp}^{(1)}(n,\mathbb{H})$ &$\xy
\POS (0,0) *\cir<3pt>{}="a",
\POS (10,0) *\cir<3pt>{} ="b",
\POS (20,0) *\cir<3pt>{} ="c",
\POS (30,0) *\cir<3pt>{} ="d",
\POS (40,0) *\cir<3pt>{} ="e",
\POS (50,0) *\cir<3pt>{} ="f"
\POS (60,0) *\cir<3pt>{} ="g"
\POS "a" \ar@{=>}^<<<{\alpha_{0}} "b",
\POS "b" \ar@{.}^<<<{\alpha_{1}} "c",
\POS "c" \ar@{-}^<<<{\alpha_{n-1}} "d",
\POS "d" \ar@{-}^<<<{\alpha_{n}} "e",
\POS "e" \ar@{.}^<<<<{\alpha_{n+1}} "f",
\POS "f" \ar@{<=}^<<<<{\alpha_{2n-1}} "g",
\POS "g" \ar@{}^<<{\alpha_{2n}},
\POS "a" \ar@{<->}@/_{1.5pc}/ "g"
\POS "b" \ar@{<->}@/_{1pc}/ "f"
\POS "c" \ar@{<->}@/_{.8pc}/ "e"
\endxy$ &$\mathfrak{sp}^{(1)}(n)$ & $\frac{SP^{(1)}(2n)}{SP^{(1)}(n)}$ &$\frac{SP^{(1)}(n,\mathbb{H})}{SP^{(1)}(n)}$ \\ \cline{2-6}
 &$\mathfrak{sp}^{(1)}(2n,\mathbb{R})\;n\geq 3$ &$\xy
\POS (0,0) *\cir<3pt>{}="a",
\POS (10,0) *\cir<3pt>{} ="b",
\POS (20,0) *\cir<3pt>{} ="c",
\POS (30,0) *{\bullet} ="d",
\POS (40,0) *\cir<3pt>{} ="e",
\POS (50,0) *\cir<3pt>{} ="f"
\POS (60,0) *\cir<3pt>{} ="g"
\POS "a" \ar@{=>}^<<<{\alpha_{0}} "b",
\POS "b" \ar@{.}^<<<{\alpha_{1}} "c",
\POS "c" \ar@{-}^<<<{\alpha_{n-1}} "d",
\POS "d" \ar@{-}^<<<{\alpha_{n}} "e",
\POS "e" \ar@{.}^<<<<{\alpha_{n+1}} "f",
\POS "f" \ar@{<=}^<<<<{\alpha_{2n-1}} "g",
\POS "g" \ar@{}^<<{\alpha_{2n}},
\POS "a" \ar@{<->}@/_{1.5pc}/ "g"
\POS "b" \ar@{<->}@/_{1pc}/ "f"
\POS "c" \ar@{<->}@/_{.8pc}/ "e"
\endxy$ & $\mathfrak{su}^{(2)}(2n)$&$\frac{SP^{(1)}(2n)}{SU^{(2)}(2n)}$ &$\frac{SP^{(1)}(2n,\mathbb{R})}{SU^{(2)}(2n)}$ \\
 \hline 
\end{tabular}
\end{sidewaystable} 

\begin{sidewaystable}[p]
\caption{Affine Kac-Moody symmetric spaces associated with $D_{n}^{(1)}$}  for even $n$ and $n>4$ 
\centering % centering table
\begin{tabular}{ | p{2cm} |p{11cm} | p{2cm}| p{3.1cm} |p{3.2cm}|}
\hline
 Real Forms&Vogan Diagram &Fixed Algebra& Compact affine Kac-Moody Symmetric spaces&Non-compact affine Kac-Moody Symmetric spaces\\
 \hline
$\mathfrak{so}^{(1)}(2n)$ & $\xy
\POS (0,0) *\cir<3pt>{} ="a",
\POS (15,0) *\cir<3pt>{} ="b",
\POS (30,0) *\cir<3pt>{} ="c",
\POS (45,0) *\cir<3pt>{} ="d",
\POS (60,0) *\cir<3pt>{} ="e",
\POS (75,5) *\cir<3pt>{} ="f",
\POS (75,-5) *\cir<3pt>{} ="g",
\POS (-15,5) *\cir<3pt>{} ="h",
\POS (-15,-5) *\cir<3pt>{} ="i",
\POS "a" \ar@{-}^<<<{\alpha_{2}} "b",
\POS "b" \ar@{.}^<<<{\alpha_{3}} "c",
\POS "c" \ar@{.}^<<{} "d",
\POS "d" \ar@{-}^<<{} "e",
\POS "e" \ar@{-}^<<{\alpha_{n-2}} "f",
\POS "g" \ar@{-}^<<{\alpha_{n}} "e",
\POS "h" \ar@{-}^<<<{\alpha_{1}} "a",
\POS "i" \ar@{-}^<<{\alpha_{0}} "a",
\POS "f" \ar@{-}^<<{\alpha_{n-1}} ,
\endxy$&$\mathfrak{so}^{(1)}(2n)$ & &\\
\hline
 &$\xy
\POS (0,0) *\cir<3pt>{} ="a",
\POS (15,0) *\cir<3pt>{} ="b",
\POS (30,0) *\cir<3pt>{} ="c",
\POS (45,0) *\cir<3pt>{} ="d",
\POS (60,0) *\cir<3pt>{} ="e",
\POS (75,5) *\cir<3pt>{} ="f",
\POS (75,-5) *\cir<3pt>{} ="g",
\POS (-15,5) *\cir<3pt>{} ="h",
\POS (-15,-5) *\cir<3pt>{} ="i",
\POS "a" \ar@{-}^<<<{\mathfrak{so}^{(1)}(4, 2n-4)} "b",
\POS "b" \ar@{.}_<<<{\mathfrak{so}^{(1)}(6, 2n-6)} "c",
\POS "c" \ar@{.}^<<{\mathfrak{so}^{(1)}(n,n)} "d",
\POS "d" \ar@{-}^<<{} "e",
\POS "e" \ar@{-}^<<{} "f",
\POS "g" \ar@{-}^<<{} "e",
\POS "h" \ar@{-}^<<<{} "a",
\POS "i" \ar@{-}^<<{\mathfrak{so}^{(1)}(2, 2n-2)} "a",
\POS "f" \ar@{-}^<<{} ,
\endxy$ &$\mathfrak{so}^{(1)}(2p)\oplus\mathfrak{so}(2q)$ &$\frac{SO^{(1)}(2p+2q)}{SO^{(1)}(2p)\oplus SO(2q)}$ &$\frac{SO^{(1)}(2p,2q)}{SO^{(1)}(2p)\oplus SO(2q)}$\\
\hline
$\mathfrak{so^{*}}^{(1)}(2n)$ & $\xy
\POS (10,-10) ="z"
\POS (0,0) *\cir<3pt>{} ="a",
\POS (15,0) *\cir<3pt>{} ="b",
\POS (30,0) *\cir<3pt>{} ="c",
\POS (45,0) *\cir<3pt>{} ="d",
\POS (60,0) *\cir<3pt>{} ="e",
\POS (75,5) *\cir<3pt>{} ="f",
\POS (75,-5) *{\bullet} ="g",
\POS (-15,5) *\cir<3pt>{} ="h",
\POS (-15,-5) *{\bullet} ="i",
\POS "a" \ar@{-}^<<<{\alpha_{2}} "b",
\POS "b" \ar@{.}^<<<{\alpha_{3}} "c",
\POS "c" \ar@{.}^<<{} "d",
\POS "d" \ar@{-}^<<{} "e",
\POS "e" \ar@{-}^<<{\alpha_{n-2}} "f",
\POS "g" \ar@{-}^<<{\alpha_{n}} "e",
\POS "h" \ar@{-}^<<<{\alpha_{1}} "a",
\POS "i" \ar@{-}^<<{\alpha_{0}} "a",
\POS "f" \ar@{-}^<<{\alpha_{n-1}} ,
\endxy$ & $\mathfrak{su}(n)$ & $\frac{SO^{(1)}(2n)}{SU(n)}$ &$\frac{{SO^{*}}^{(1)}(n)}{SU(n)}$ \\
\hline
$\mathfrak{so}_{-1}^{(1)}(2, 2n-2)$ & $\xy
\POS (10,-10) ="z"
\POS (0,0) *\cir<3pt>{} ="a",
\POS (15,0) *\cir<3pt>{} ="b",
\POS (30,0) *\cir<3pt>{} ="c",
\POS (45,0) *\cir<3pt>{} ="d",
\POS (60,0) *\cir<3pt>{} ="e",
\POS (75,5) *\cir<3pt>{} ="f",
\POS (75,-5) *\cir<3pt>{} ="g",
\POS (-15,5) *{\bullet}="h",
\POS (-15,-5) *{\bullet} ="i",
\POS "a" \ar@{-}^<<<{\alpha_{2}} "b",
\POS "b" \ar@{.}^<<<{\alpha_{3}} "c",
\POS "c" \ar@{.}^<<{} "d",
\POS "d" \ar@{-}^<<{} "e",
\POS "e" \ar@{-}^<<{\alpha_{n-2}} "f",
\POS "g" \ar@{-}^<<{\alpha_{n}} "e",
\POS "h" \ar@{-}^<<<{\alpha_{1}} "a",
\POS "i" \ar@{-}^<<{\alpha_{0}} "a",
\POS "f" \ar@{-}^<<{\alpha_{n-1}} ,
\endxy$ &$\mathfrak{so}(2n-2)$ & $\frac{SO^{(1)}(2n)}{SO(2n-2)}$ &$\frac{SO_{-1}^{(1)}(2,2n-2)}{SO(2n-2)}$\\
\hline 
$\mathfrak{so}_{\sigma_{v}}^{(1)}(1, 2n-1)$&$\xy
\POS (10,-10) ="z"
\POS (0,0) *\cir<3pt>{} ="a",
\POS (15,0) *\cir<3pt>{} ="b",
\POS (30,0) *\cir<3pt>{} ="c",
\POS (45,0) *\cir<3pt>{} ="d",
\POS (60,0) *\cir<3pt>{} ="e",
\POS (75,5) *\cir<3pt>{} ="f",
\POS (75,-5) *\cir<3pt>{} ="g",
\POS (-15,5) *\cir<3pt>{} ="h",
\POS (-15,-5) *\cir<3pt>{} ="i",
\POS "a" \ar@{-}^<<<{\alpha_{2}} "b",
\POS "b" \ar@{.}^<<<{\alpha_{3}} "c",
\POS "c" \ar@{.}^<<{} "d",
\POS "d" \ar@{-}^<<{} "e",
\POS "e" \ar@{-}^<<{\alpha_{n-2}} "f",
\POS "g" \ar@{-}^<<{\alpha_{n}} "e",
\POS "h" \ar@{-}^<<<{\alpha_{1}} "a",
\POS "i" \ar@{-}^<<{\alpha_{0}} "a",
\POS "f" \ar@{-}^<<{\alpha_{n-1}} ,
\POS "i" \ar@{<->}@/^{1.5pc}/ "h",
\POS "g" \ar@{<->}@/_{1.5pc}/ "f",
\endxy$&$\mathfrak{sp}^{(2)}(2n-2)$& $\frac{SO^{(1)}(2n)}{SP^{(2)}(2n-2)}$&$\frac{SO_{\sigma_{v}}^{(1)}(1, 2n-1)}{SP^{(2)}(2n-2)}$\\
\hline
&$\xy
\POS (10,-10) ="z"
\POS (0,0) *\cir<3pt>{} ="a",
\POS (15,0) *\cir<3pt>{} ="b",
\POS (30,0) *\cir<3pt>{} ="c",
\POS (45,0) *\cir<3pt>{} ="d",
\POS (60,0) *\cir<3pt>{} ="e",
\POS (75,5) *\cir<3pt>{} ="f",
\POS (75,-5) *\cir<3pt>{} ="g",
\POS (-15,5) *\cir<3pt>{} ="h",
\POS (-15,-5) *\cir<3pt>{} ="i",
\POS "a" \ar@{-}^<<<{\mathfrak{so}_{\sigma_{v}}^{(1)}(5, 2n-5)} "b",
\POS "b" \ar@{.}_<<<{\mathfrak{so}_{\sigma_{v}}^{(1)}(7, 2n-7)} "c",
\POS "c" \ar@{.}^<<{\mathfrak{so}_{\sigma_{v}}^{(1)}(n+1, n-1)} "d",
\POS "d" \ar@{-}^<<{} "e",
\POS "e" \ar@{-}^<<{} "f",
\POS "g" \ar@{-}^<<{} "e",
\POS "h" \ar@{-}^<<<{} "a",
\POS "i" \ar@{-}^<<{} "a",
\POS "f" \ar@{-}^<<{},
\POS "i" \ar@{<->}@/^{1.5pc}/ "h",
\POS "g" \ar@{<->}@/_{1.5pc}/ "f",
\endxy$ &$\mathfrak{so}^{(2)}(2p)\oplus\mathfrak{so}(2q+1)$& $\frac{SO^{(1)}(2n)}{SO^{(2)}(2p)\oplus SO(2q+1)}$ &$\frac{SO_{\sigma_{v}}^{(1)}(2p, 2q+1)}{SO^{(2)}(2p)\oplus SO(2q+1)}$\\
\hline
$\mathfrak{so}_{\gamma}^{(1)}(1, 2n-1)$&$\xy
\POS (10,-10) ="z"
\POS (0,0) *\cir<3pt>{} ="a",
\POS (15,0) *\cir<3pt>{} ="b",
\POS (30,0) *\cir<3pt>{} ="c",
\POS (45,0) *\cir<3pt>{} ="d",
\POS (60,0) *\cir<3pt>{} ="e",
\POS (75,8) *\cir<3pt>{} ="f",
\POS (75,-8) *\cir<3pt>{} ="g",
\POS (-15,8) *\cir<3pt>{} ="h",
\POS (-15,-8) *\cir<3pt>{} ="i",
\POS "a" \ar@{-}^<<<{\alpha_{2}} "b",
\POS "b" \ar@{.}^<<<{\alpha_{3}} "c",
\POS "c" \ar@{.}^<<{} "d",
\POS "d" \ar@{-}^<<{} "e",
\POS "e" \ar@{-}^<<{\alpha_{n-2}} "f",
\POS "g" \ar@{-}^<<{\alpha_{n}} "e",
\POS "h" \ar@{-}^<<<{\alpha_{1}} "a",
\POS "i" \ar@{-}^<<{\alpha_{0}} "a",
\POS "f" \ar@{-}^<<{\alpha_{n-1}} ,
\POS "g" \ar@{<->}@/_{1.5pc}/ "f",
\endxy$&$\mathfrak{so}^{(1)}(2n-1)$ &$\frac{SO^{(1)}(2n)}{SO^{(1)}(2n-1)}$ &$\frac{SO_{\gamma}^{(1)}(1, 2n-1)}{SO^{(1)}(2n-1)}$\\
\hline
&$\xy
\POS (0,0) *\cir<3pt>{} ="a",
\POS (15,0) *\cir<3pt>{} ="b",
\POS (30,0) *\cir<3pt>{}="c",
\POS (45,0) *\cir<3pt>{} ="d",
\POS (60,0) *\cir<3pt>{} ="e",
\POS (75,8) *\cir<3pt>{} ="f",
\POS (75,-8) *\cir<3pt>{} ="g",
\POS (-15,8) *\cir<3pt>{} ="h",
\POS (-15,-8) *\cir<3pt>{} ="i",
\POS "a" \ar@{-}^<<<{\mathfrak{so}_{\gamma}^{(1)}(5, 2n-5)} "b",
\POS "b" \ar@{.}_<<<{\mathfrak{so}_{\gamma}^{(1)}(7, 2n-7)} "c",
\POS "c" \ar@{.}^<<{\mathfrak{so}_{\gamma}^{(1)}(n+1, n-1)} "d",
\POS "d" \ar@{-}^<<{} "e",
\POS "e" \ar@{-}^<<{\mathfrak{so}_{\gamma}^{(1)}(2n-5,5)} "f",
\POS "g" \ar@{-}^<<{} "e",
\POS "h" \ar@{-}^<<<{} "a",
\POS "i" \ar@{-}^<<{\mathfrak{so}_{\gamma}^{(1)}(3, 2n-3)} "a",
\POS "f" \ar@{-}^<<{} ,
\POS "g" \ar@{<->}@/_{1.5pc}/ "f",
\endxy$ & $\mathfrak{so}^{(1)}(2p+1)\oplus\mathfrak{so}(2q+1)$& $\frac{SO^{(1)}(2n)}{SO^{(1)}(2p+1)\oplus SO(2q+1)}$ &$\frac{SO_{\gamma}^{(1)}(2p+1, 2q+1}{SO^{(1)}(2p+1)\oplus SO(2q+1)}$\\
\hline
$\mathfrak{so}_{\sigma_{s}}^{(1)}(1, 2n-1)$&$\xy
\POS (10,-10) ="z"
\POS (0,0) *\cir<3pt>{} ="a",
\POS (15,0) *\cir<3pt>{} ="b",
\POS (30,0) *\cir<3pt>{} ="c",
\POS (45,0) *\cir<3pt>{} ="d",
\POS (60,0) *\cir<3pt>{} ="e",
\POS (75,7) *\cir<3pt>{} ="f",
\POS (75,-7) *\cir<3pt>{} ="g",
\POS (-15,7) *\cir<3pt>{} ="h",
\POS (-15,-7) *\cir<3pt>{} ="i",
\POS "a" \ar@{-}^<<<{\alpha_{2}} "b",
\POS "b" \ar@{.}^<<<{\alpha_{3}} "c",
\POS "c" \ar@{.}^<<{} "d",
\POS "d" \ar@{-}^<<{} "e",
\POS "e" \ar@{-}^<<{\alpha_{n-2}} "f",
\POS "g" \ar@{-}^<<{\alpha_{n}} "e",
\POS "h" \ar@{-}^<<<{\alpha_{1}} "a",
\POS "i" \ar@{-}^<<{\alpha_{0}} "a",
\POS "f" \ar@{-}^<<{\alpha_{n-1}} ,
\POS "f" \ar@{<->}^<<{} "h",
\POS "i" \ar@{<->}^<<{}"g",
\POS "a" \ar@{<->}@/^{1.5pc}/ "e",
\POS "b" \ar@{<->}@/_{1.5pc}/ "d",
\endxy$&$\mathfrak{su}^{(2)}(n)$ &$\frac{SO^{(1)}(2n)}{SU^{(2)}(n)}$  &$\frac{SO_{\sigma_{s}}^{(1)}(1, 2n-1)}{SU^{(2)}(n)}$\\
\hline
$\mathfrak{so}_{\sigma_{s}}^{(1)}(n+1, n-1)$&$\xy
\POS (10,-10) ="z"
\POS (0,0) *\cir<3pt>{} ="a",
\POS (15,0) *\cir<3pt>{} ="b",
\POS (30,0) *{\bullet} ="c",
\POS (45,0) *\cir<3pt>{} ="d",
\POS (60,0) *\cir<3pt>{} ="e",
\POS (75,7) *\cir<3pt>{} ="f",
\POS (75,-7) *\cir<3pt>{} ="g",
\POS (-15,7) *\cir<3pt>{} ="h",
\POS (-15,-7) *\cir<3pt>{} ="i",
\POS "a" \ar@{-}^<<<{\alpha_{2}} "b",
\POS "b" \ar@{.}^<<<{\alpha_{3}} "c",
\POS "c" \ar@{.}^<<{} "d",
\POS "d" \ar@{-}^<<{} "e",
\POS "e" \ar@{-}^<<{\alpha_{n-2}} "f",
\POS "g" \ar@{-}^<<{\alpha_{n}} "e",
\POS "h" \ar@{-}^<<<{\alpha_{1}} "a",
\POS "i" \ar@{-}^<<{\alpha_{0}} "a",
\POS "f" \ar@{-}^<<{\alpha_{n-1}} ,
\POS "f" \ar@{<->}^<<{} "h",
\POS "i" \ar@{<->}^<<{}"g",
\POS "a" \ar@{<->}@/^{1.5pc}/ "e",
\POS "b" \ar@{<->}@/_{1.5pc}/ "d",
\endxy$&$\mathfrak{so}^{(1)(n)}$ & $\frac{SO^{(1)(2n)}}{SO^{(1)(n)}}$ &$\frac{SO_{\sigma_{s}}^{(1)}(n+1, n-1)}{SO^{(1)(n)}}$\\
\hline
\end{tabular}
\end{sidewaystable}  

\begin{sidewaystable}[p]
\caption{Affine Kac-Moody symmetric spaces associated with $D_{n}^{(1)}$}  for $n$ odd and $n>4$ 
\centering % centering table
\begin{tabular}{ | p{2cm} |p{11cm} | p{2cm}| p{3.1cm} |p{3.2cm}|}
\hline
 Real Forms&Vogan Diagram &Fixed Algebra& Compact affine Kac-Moody Symmetric spaces&Non-compact affine Kac-Moody Symmetric spaces\\
 \hline
$\mathfrak{so}^{(1)}(2n)$ &$\xy
\POS (0,0) *\cir<3pt>{} ="a",
\POS (15,0) *\cir<3pt>{} ="b",
\POS (30,0) *\cir<3pt>{} ="c",
\POS (45,0) *\cir<3pt>{} ="d",
\POS (60,5) *\cir<3pt>{} ="e",
\POS (60,-5) *\cir<3pt>{} ="f",
\POS (-15,5) *\cir<3pt>{} ="g",
\POS (-15,-5) *\cir<3pt>{} ="h",
\POS "a" \ar@{.}^<<<{\alpha_{2}} "b",
\POS "b" \ar@{-}^<<<{} "c",
\POS "c" \ar@{.}^<<{} "d",
\POS "e" \ar@{-}^<<{\alpha_{n-1}} "d",
\POS "d" \ar@{-}^<<{\alpha_{n-2}} "f",
\POS "g" \ar@{-}^<<{\alpha_{1}} "a",
\POS "h" \ar@{-}^<<<{\alpha_{0}} "a",
\POS "f" \ar@{-}^<<{\alpha_{n}} 
\endxy$ & $\mathfrak{so}^{(1)}(2n)$ & &\\
\hline
 &$\xy
\POS (0,0) *\cir<3pt>{} ="a",
\POS (15,0) *\cir<3pt>{} ="b",
\POS (30,0) *\cir<3pt>{} ="c",
\POS (45,0) *\cir<3pt>{} ="d",
\POS (60,5) *\cir<3pt>{} ="e",
\POS (60,-5) *\cir<3pt>{} ="f",
\POS (-15,5) *\cir<3pt>{} ="g",
\POS (-15,-5) *\cir<3pt>{} ="h",
\POS "a" \ar@{.}^<<<{\mathfrak{so}^{(1)}(4, 2n-4)} "b",
\POS "b" \ar@{-}_<<<{\mathfrak{so}^{(1)}(n-1, n+1)} "c",
\POS "c" \ar@{.}^<<{\mathfrak{so}^{(1)}(n+1, n-1)} "d",
\POS "e" \ar@{-}^<<{} "d",
\POS "d" \ar@{-}^<<{} "f",
\POS "g" \ar@{-}^<<{} "a",
\POS "h" \ar@{-}^<<<{\mathfrak{so}^{(1)}(2, 2n-2)} "a",
\POS "f" \ar@{-}^<<{} 
\endxy$ &$\mathfrak{so}^{(1)}(2p)\oplus\mathfrak{so}(2q)$ &$\frac{SO^{(1)}(2p+2q)}{SO^{(1)}(2p)\oplus SO(2q)}$ &$\frac{SO^{(1)}(2p,2q)}{SO^{(1)}(2p)\oplus SO(2q)}$\\
\hline
$\mathfrak{so^{*}}^{(1)}(2n)$ &$\xy
\POS (0,0) *\cir<3pt>{} ="a",
\POS (15,0) *\cir<3pt>{} ="b",
\POS (30,0) *\cir<3pt>{} ="c",
\POS (45,0) *\cir<3pt>{} ="d",
\POS (60,10) *\cir<3pt>{} ="e",
\POS (60,-10) *{\bullet} ="f",
\POS (-15,10) *\cir<3pt>{} ="g",
\POS (-15,-10) *{\bullet} ="h",
\POS "a" \ar@{.}^<<<{\alpha_{2}} "b",
\POS "b" \ar@{-}^<<<{} "c",
\POS "c" \ar@{.}^<<{} "d",
\POS "e" \ar@{-}^<<{\alpha_{n-1}} "d",
\POS "d" \ar@{-}^<<{\alpha_{n-2}} "f",
\POS "g" \ar@{-}^<<{\alpha_{1}} "a",
\POS "h" \ar@{-}^<<<{\alpha_{0}} "a",
\POS "f" \ar@{-}^<<{\alpha_{n}} 
\endxy$ &$\mathfrak{su}(n)$ & $\frac{SO^{(1)}(2n)}{SU(n)}$ &$\frac{{SO^{*}}^{(1)}(n)}{SU(n)}$\\
\hline
$\mathfrak{so}_{-1}^{(1)}(2, 2n-2)$ &$\xy
\POS (0,0) *\cir<3pt>{} ="a",
\POS (15,0) *\cir<3pt>{} ="b",
\POS (30,0) *\cir<3pt>{} ="c",
\POS (45,0) *\cir<3pt>{} ="d",
\POS (60,10) *\cir<3pt>{} ="e",
\POS (60,-10) *\cir<3pt>{} ="f",
\POS (-15,10) *{\bullet} ="g",
\POS (-15,-10) *{\bullet} ="h",
\POS "a" \ar@{.}^<<<{\alpha_{2}} "b",
\POS "b" \ar@{-}^<<<{} "c",
\POS "c" \ar@{.}^<<{} "d",
\POS "e" \ar@{-}^<<{\alpha_{n-1}} "d",
\POS "d" \ar@{-}^<<{\alpha_{n-2}} "f",
\POS "g" \ar@{-}^<<{\alpha_{1}} "a",
\POS "h" \ar@{-}^<<<{\alpha_{0}} "a",
\POS "f" \ar@{-}^<<{\alpha_{n}} 
\endxy$ &$\mathfrak{so}(2n-2)$ & $\frac{SO^{(1)}(2n)}{SO(2n-2)}$ &$\frac{SO_{-1}^{(1)}(2,2n-2)}{SO(2n-2)}$\\
\hline
$\mathfrak{so}_{\sigma_{v}}^{(1)}(1, 2n-1)$ &$\xy
\POS (0,0) *\cir<3pt>{} ="a",
\POS (15,0) *\cir<3pt>{} ="b",
\POS (30,0) *\cir<3pt>{} ="c",
\POS (45,0) *\cir<3pt>{} ="d",
\POS (60,8) *\cir<3pt>{} ="e",
\POS (60,-8) *\cir<3pt>{} ="f",
\POS (-15,8) *\cir<3pt>{} ="g",
\POS (-15,-8) *\cir<3pt>{} ="h",
\POS "a" \ar@{.}^<<<{\alpha_{2}} "b",
\POS "b" \ar@{-}^<<<{} "c",
\POS "c" \ar@{.}^<<{} "d",
\POS "e" \ar@{-}^<<{\alpha_{n-1}} "d",
\POS "d" \ar@{-}^<<{\alpha_{n-2}} "f",
\POS "g" \ar@{-}^<<{\alpha_{1}} "a",
\POS "h" \ar@{-}^<<<{\alpha_{0}} "a",
\POS "f" \ar@{-}^<<{\alpha_{n}},
\POS "e" \ar@{<->}@/^{1.5pc}/ "f",
\POS "g" \ar@{<->}@/_{1.5pc}/ "h",  
\endxy$ &$\mathfrak{sp}^{(2)}(2n-2)$ &$\frac{SO^{(1)}(2n)}{SP^{(2)}(2n-2)} $&$\frac{SO_{\sigma_{v}}^{(1)}(1, 2n-1)}{SP^{(2)}(2n-2)} $\\
\hline
&$\xy
\POS (10,-10) ="z"
\POS (0,0) *\cir<3pt>{} ="a",
\POS (15,0) *\cir<3pt>{} ="b",
\POS (30,0) *\cir<3pt>{} ="c",
\POS (45,0) *\cir<3pt>{} ="d",
\POS (60,0) *\cir<3pt>{} ="e",
\POS (75,5) *\cir<3pt>{} ="f",
\POS (75,-5) *\cir<3pt>{} ="g",
\POS (-15,5) *\cir<3pt>{} ="h",
\POS (-15,-5) *\cir<3pt>{} ="i",
\POS "a" \ar@{-}^<<<{\mathfrak{so}_{\sigma_{v}}^{(1)}(5, 2n-5)} "b",
\POS "b" \ar@{.}_<<<{\mathfrak{so}_{\sigma_{v}}^{(1)}(7, 2n-7)} "c",
\POS "c" \ar@{.}^<<{\mathfrak{so}_{\sigma_{v}}^{(1)}(n+1, n-1)} "d",
\POS "d" \ar@{-}^<<{} "e",
\POS "e" \ar@{-}^<<{} "f",
\POS "g" \ar@{-}^<<{} "e",
\POS "h" \ar@{-}^<<<{} "a",
\POS "i" \ar@{-}^<<{} "a",
\POS "f" \ar@{-}^<<{},
\POS "i" \ar@{<->}@/^{1.5pc}/ "h",
\POS "g" \ar@{<->}@/_{1.5pc}/ "f",
\endxy$ &$\mathfrak{so}^{(2)}(2p)\oplus\mathfrak{so}(2q+1)$& $\frac{SO^{(1)}(2n)}{SO^{(2)}(2p)\oplus SO(2q+1)}$ &$\frac{SO^{(1)}(2p, 2q+1)}{SO^{(2)}(2p)\oplus SO(2q+1)}$\\ \hline
$\mathfrak{so}_{\gamma}^{(1)}(1, 2n-1)$ &$\xy
\POS (0,0) *\cir<3pt>{} ="a",
\POS (15,0) *\cir<3pt>{} ="b",
\POS (30,0) *\cir<3pt>{} ="c",
\POS (45,0) *\cir<3pt>{} ="d",
\POS (60,8) *\cir<3pt>{} ="e",
\POS (60,-8) *\cir<3pt>{} ="f",
\POS (-15,8) *\cir<3pt>{} ="g",
\POS (-15,-8) *\cir<3pt>{} ="h",
\POS "a" \ar@{.}^<<<{} "b",
\POS "b" \ar@{-}_<<<{} "c",
\POS "c" \ar@{.}^<<{} "d",
\POS "e" \ar@{-}^<<{} "d",
\POS "d" \ar@{-}^<<{} "f",
\POS "g" \ar@{-}^<<{} "a",
\POS "h" \ar@{-}^<<<{} "a",
\POS "f" \ar@{-}^<<{},
\POS "e" \ar@{<->}@/^{1.5pc}/ "f", 
\endxy$ & $\mathfrak{so}^{(1)}(2n-1)$ &$\frac{SO^{(1)}(2n)}{SO^{(1)}(2n-1)} $&$\frac{SO_{\gamma}^{(1)}(1, 2n-1)}{SO^{(1)}(2n-1)} $\\
\hline
&$\xy
\POS (0,0) *\cir<3pt>{} ="a",
\POS (15,0) *\cir<3pt>{} ="b",
\POS (30,0) *\cir<3pt>{}="c",
\POS (45,0) *\cir<3pt>{} ="d",
\POS (60,0) *\cir<3pt>{} ="e",
\POS (75,8) *\cir<3pt>{} ="f",
\POS (75,-8) *\cir<3pt>{} ="g",
\POS (-15,8) *\cir<3pt>{} ="h",
\POS (-15,-8) *\cir<3pt>{} ="i",
\POS "a" \ar@{-}^<<<{\mathfrak{so}_{\gamma}^{(1)}(5, 2n-5)} "b",
\POS "b" \ar@{.}_<<<{\mathfrak{so}_{\gamma}^{(1)}(7, 2n-7)} "c",
\POS "c" \ar@{.}^<<{\mathfrak{so}_{\gamma}^{(1)}(n+1, n-1)} "d",
\POS "d" \ar@{-}^<<{} "e",
\POS "e" \ar@{-}^<<{\mathfrak{so}_{\gamma}^{(1)}(2n-5,5)} "f",
\POS "g" \ar@{-}^<<{} "e",
\POS "h" \ar@{-}^<<<{} "a",
\POS "i" \ar@{-}^<<{\mathfrak{so}_{\gamma}^{(1)}(3, 2n-3)} "a",
\POS "f" \ar@{-}^<<{} ,
\POS "g" \ar@{<->}@/_{1.5pc}/ "f",
\endxy$ &$\mathfrak{so}^{(1)}(2p+1)\oplus\mathfrak{so}(2q+1)$ &$\frac{SO^{(1)}(2n)}{SO^{(1)}(2p+1)\oplus SO(2q+1)}$ &$\frac{SO^{(1)}(2p+1, 2q+1)}{SO^{(1)}(2p+1)\oplus SO(2q+1)}$\\
\hline
$\mathfrak{so}_{\sigma_{s}}^{(1)}(1, 2n-1)$&$\xy
\POS (0,0) *\cir<3pt>{} ="a",
\POS (15,0) *\cir<3pt>{} ="b",
\POS (30,0) *\cir<3pt>{} ="c",
\POS (45,0) *\cir<3pt>{} ="d",
\POS (60,12) *\cir<3pt>{} ="e",
\POS (60,-12) *\cir<3pt>{} ="f",
\POS (-15,12) *\cir<3pt>{} ="g",
\POS (-15,-12) *\cir<3pt>{} ="h",
\POS "a" \ar@{.}^<<<{\alpha_{2}} "b",
\POS "b" \ar@{-}^<<<{} "c",
\POS "c" \ar@{.}^<<{} "d",
\POS "e" \ar@{-}^<<{\alpha_{n-1}} "d",
\POS "d" \ar@{-}^<<{\alpha_{n-2}} "f",
\POS "g" \ar@{-}^<<{\alpha_{1}} "a",
\POS "h" \ar@{-}^<<<{\alpha_{0}} "a",
\POS "f" \ar@{-}^<<{\alpha_{n}},
\POS "h" \ar@{->}^<<{} "f",
\POS "f" \ar@{->}^<<{} "g",
\POS "g" \ar@{->}^<<{} "e",
\POS "a" \ar@{<->}@/^{1.5pc}/ "d",
\POS "b" \ar@{<->}@/^{1.2pc}/ "c",  
\endxy$ & $\mathfrak{so}^{(2)}(n-1)$  & $\frac{SO_{\sigma_{s}}^{(1)}(1, 2n-1)}{SO^{(2)}(n-1)} $&$\frac{SO_{\sigma_{s}}^{(1)}(1, 2n-1)}{SO^{(2)}(n-1)} $\\
\hline
\end{tabular}
\end{sidewaystable} 

\newpage
\section{Appendix}
\subsection{Classical Irreducible Reduced Root systems}
In this subsection we have given the irreducible reduced root system of complex semi simple Lie algebras $A_{n}$ for $n\geq 1$, $B_{n}$ for $n\geq 2$, $C_{n}$ for $n\geq 3$ and $D_{n}$ for $n\geq 4$. In case of $A_{n}$ the under lying vector space $V=\{v\in \mathbb{R}^{n+1}\mid <v, e_{1}+\cdots+e_{n+1}>=0\}$ and for rest algebras $V=\mathbb{R}^{n}. \Delta$ denotes the root system which is a subspace of some $\mathbb{R}^{k}=\sum_{i=1}^{k}a_{i}e_{i}$. Here $\{e_{i}\}$ is the standard orthonormal basis and $a_{i}$'s are real. $\Delta^{+}$ is the positive root system and $\Pi$ is simple root system.
\begin{table}[h]
\caption{}
\centering % centering table
\begin{tabular}{ | l| p{2.7cm} |p{2.7cm} | p{4cm}| p{2cm} |}
\hline
 $g$ & $\Delta$ &$\Delta^{+}$& $\Pi$&Largest Root\\ \hline
 $A_{n}=\mathfrak{sl}(n+1,\mathbb{C})$&$\{e_{i}-e_{j}\mid i\neq j\}$&$\{e_{i}-e_{j}\mid i<j\}$& $\{e_{1}-e_{2},\cdots, e_{n}-e_{n+1}\}$&$e_{1}-e_{n+1}$\\ \hline
 $B_{n}=\mathfrak{so}(2n+1,\mathbb{C})$&$\{\pm e_{i}\pm e_{j}\mid i< j\}\cup \{\pm e_{i}\}$&$\{e_{i}\pm e_{j}\mid i<j\}\cup\{e_{i}\}$& $\{e_{1}-e_{2},\cdots, e_{n-1}-e_{n}, e_{n}\}$&$e_{1}+e_{2}$\\ \hline
 $C_{n}=\mathfrak{sp}(n,\mathbb{C})$&$\{\pm e_{i}\pm e_{j}\mid i< j\}\cup \{\pm 2 e_{i}\}$&$\{e_{i}\pm e_{j}\mid i<j\}\cup\{2e_{i}\}$& $\{e_{1}-e_{2},\cdots, e_{n-1}-e_{n}, 2e_{n}\}$&$2e_{1}$\\ \hline
 $D_{n}=\mathfrak{so}(2n,\mathbb{C})$&$\{\pm e_{i}\pm e_{j}\mid i< j\}$&$\{e_{i}\pm e_{j}\mid i<j\}$& $\{e_{1}-e_{2},\cdots, e_{n-1}-e_{n}, e_{n-1}+e_{n}\}$&$e_{1}+e_{2}$\\ \hline
 
\end{tabular}
\end{table}

\subsection{Diagram Automorphism}
The diagram automorphism for affine Lie algebra are as follows. For $A_{n}^{(1)}$, the automorphism group of the Dynkin diagram is the dihedral group $D_{n+1}$ which is generated by reflection $s:i\longmapsto n+1-i(\textrm{mod}\ n+1)$ and the rotation $r:i\longmapsto i+1(\textrm{mod}\ n+1)$ which is of order $r+1$. For $D_{r}^{(1)}$, the automorphism group is generated by the vector automorphism $\sigma_{v}$, the spinor automorphism $\sigma_{s}$ and the conjugation $\gamma$. $\sigma_{v}$ acts as $0\longleftrightarrow 1,\;r\longleftrightarrow r-1$ and $i\longmapsto i$ else, and hence is of order 2. The map $\gamma$ acts as  $r\longleftrightarrow r-1$ and $i\longmapsto i$ else. If $r$ is even, $\sigma_{s}$ acts as $i\longmapsto r-i$ is of order 2 while for odd $r$ the prescription $i\longmapsto r-i$ only holds for $2\leq i\leq r-2$ and is supplemented by $0\longmapsto r\longmapsto 1\longmapsto r-1\longmapsto 0$. For the untwisted algebra $g=B_{r}^{(1)},C_{r}^{(1)} $ and for twisted algebra $g=B_{r}^{(2)},C_{r}^{(2)}$, there is only a single non-trivial automorphism $\gamma$ which is a reflection \cite{Fuchs1996}.
\subsection{Classical Non-compact real affine Kac-Moody Lie Algebras}
Let $g_{\mathbb{R}}$ be a real affine Kac-Moody Lie algebra and $\mathfrak{t}_{0}$ be the fixed subalgebra of a Cartan involution on $g_{\mathbb{R}}$. $\mathfrak{c_{0}}$ is the center of $\mathfrak{t}_{0}$, then  simple roots of $\mathfrak{t_{0}}$ are obtained as follows. When the automorphism in the Vogan diagram is non trivial, we know that $\mathfrak{t_{0}}$ is semisimple. The simple roots for $\mathfrak{t_{0}}$ then include the compact imaginary simple roots and the average of the members of each 2-element orbit of simple roots. If the Vogan diagram has no painted imaginary root, there is no other simple root for $\mathfrak{t_{0}}$. Otherwise there is one other simple root for $\mathfrak{t_{0}}$, obtained by taking a minimal complex root containing the painted imaginary root in its expansion and averaging it over its 2-element orbit under the automorphism. When the automorphism is trivial, either $\dim\mathfrak{c_{0}}=1$, in this case the simple roots for $\mathfrak{t_{0}}$ are the compact simple roots for $\mathfrak{g_{0}}$, or else  $\dim\mathfrak{c_{0}}=0$, in this case the simple roots for $\mathfrak{t_{0}}$ are the compact imaginary simple roots for $\mathfrak{g_{0}}$ and one other compact imaginary root. In latter case this other compact imaginary root is the unique smallest root containing the non-compact simple root twice in its expansion. We have discussed below in detail, for each real algebra separately.
\begin{itemize}
 \item $\mathfrak{sl}_{s}^{(1)}(n, \mathbb{H}), n\;\; \mathrm{even} \geq 2$\\
Vogan diagram:\\ 
\hspace*{2.5cm}$A_{n-1}$, non trivial automorphism,\\
\hspace*{2.5cm} no imaginary simple roots\\
 $\mathfrak{t_{0}}=\mathfrak{sp}^{(1)}(n)$\\
 Simple roots for $\mathfrak{t_{0}}$:\\
 \hspace*{2.5cm} $e_{2n}-e_{1},\;e_{n}-e_{n+1}$ and \\
 \hspace*{2.5cm}all $\frac{1}{2}(e_{i}-e_{i+1}+e_{2n-i}-e_{2n+1-i})$ for $1\leq i \leq (n-1)$

\item $\mathfrak{sl}_{-1}^{(1)}(2n, \mathbb{R}), n \geq 3$\\
Vogan diagram:\\ 
\hspace*{2.5cm}$A_{n-1}$, non trivial automorphism,\\
\hspace*{2.5cm} unique imaginary simple root $e_{2n}-e_{1}$\\
 $\mathfrak{t_{0}}=\mathfrak{su}^{(2)}(2n)$\\
 Simple roots for $\mathfrak{t_{0}}$:\\
 \hspace*{2.5cm} $\frac{1}{2}(e_{n-1}+e_{n}-e_{n+1}-e_{n+2})$ and \\
 \hspace*{2.5cm}all $\frac{1}{2}(e_{i}-e_{i+1}+e_{2n-i}-e_{2n+1-i})$ for $1\leq i \leq (n-1)$
 
\item $\mathfrak{sl}_{-1}^{(1)}(2n+1, \mathbb{R}), n \geq 3$\\
Vogan diagram:\\ 
\hspace*{2.5cm}$A_{n-1}$, non trivial automorphism,\\
\hspace*{2.5cm} no imaginary simple roots \\
 $\mathfrak{t_{0}}=\mathfrak{su}^{(2)}(2n+1)$\\
 Simple roots for $\mathfrak{t_{0}}$:\\
 \hspace*{2.5cm} $e_{2n+1}-e_{1},\; \frac{1}{2}(e_{n}-e_{n+2})$ and \\
 \hspace*{2.5cm}all $\frac{1}{2}(e_{i}-e_{i+1}+e_{2n+1-i}-e_{2n+2-i})$ for $1\leq i \leq (n-1)$

\item $\mathfrak{sl}_{1}^{(1)}(2n, \mathbb{R}), n \geq 4$\\
Vogan diagram:\\ 
\hspace*{2.5cm}$A_{n-1}$, non trivial automorphism,\\
\hspace*{2.5cm} two imaginary simple roots $e_{2n}-e_{1},e_{n}-e_{n+1}$\\
 $\mathfrak{t_{0}}=\mathfrak{so}^{(1)}(2n)$\\
 Simple roots for $\mathfrak{t_{0}}$:\\
 \hspace*{2.5cm} $\frac{1}{2}(e_{n-1}+e_{n}-e_{n+1}-e_{n+2}),\frac{1}{2}(e_{2n-1}+e_{2n}-e_{1}-e_{2}) $ and \\
 \hspace*{2.5cm}all $\frac{1}{2}(e_{i}-e_{i+1}+e_{2n+1-i}-e_{2n+2-i})$ for $1\leq i \leq (n-1)$
 
\item $\mathfrak{sl}_{1}^{(1)}(2n+1, \mathbb{R}), n \geq 4$\\
Vogan diagram:\\ 
\hspace*{2.5cm}$A_{n-1}$, non trivial automorphism,\\
\hspace*{2.5cm} one imaginary simple roots $e_{2n+1}-e_{1}$\\
 $\mathfrak{t_{0}}=\mathfrak{so}^{(1)}(2n)$\\
 Simple roots for $\mathfrak{t_{0}}$:\\
 \hspace*{2.5cm} $\frac{1}{2}(e_{n}-e_{n+2}),\frac{1}{2}(e_{2n}+e_{2n+1}-e_{1}-e_{2}) $ and \\
 \hspace*{2.5cm}all $\frac{1}{2}(e_{i}-e_{i+1}+e_{2n+1-i}-e_{2n+2-i})$ for $1\leq i \leq (n-1)$

\item $\mathfrak{sl}_{r^{n}}^{(1)}(2n, \mathbb{H}), n \geq 4$\\
Vogan diagram:\\ 
\hspace*{2.5cm}$A_{n-1}$, non trivial automorphism,\\
\hspace*{2.5cm} no imaginary simple roots \\
 $\mathfrak{t_{0}}=\mathfrak{su}^{(1)}(n)$\\
 Simple roots for $\mathfrak{t_{0}}$:\\
 \hspace*{2.5cm} $\frac{1}{2}(e_{n}-e_{n+1}+e_{2n}-e_{1})$ and \\
 \hspace*{2.5cm}all $\frac{1}{2}(e_{i}-e_{i+1}+e_{n+i}-e_{n+1-i})$ for $1\leq i \leq (n-1)$

\item $\mathfrak{sl}_{rs}^{(1)}(n, \mathbb{H}), n \geq 4$\\
Vogan diagram:\\ 
\hspace*{2.5cm}$A_{n-1}$, non trivial automorphism,\\
\hspace*{2.5cm} no imaginary simple roots \\
 $\mathfrak{t_{0}}=\mathfrak{so}^{(2)}(2n)$\\
 Simple roots for $\mathfrak{t_{0}}$:\\
 \hspace*{2.5cm} $\frac{1}{2}(e_{2n}-e_{2}),\;\frac{1}{2}(e_{n}-e_{n+2})$ and \\
 \hspace*{2.5cm}all $\frac{1}{2}(e_{i+1}-e_{i+2}+e_{2n-i}-e_{2n+1-i})$ for $1\leq i \leq (n-2)$

\item $\mathfrak{su}_{-1}^{(1)}(p, q), p+q=2n$\\
Vogan diagram:\\ 
\hspace*{2.5cm}$A_{2n-1}$, trivial automorphism,\\
\hspace*{2.5cm} unique imaginary simple root $e_{2n}-e_{1}$ \\
 $\mathfrak{t_{0}}=\mathfrak{su}(2n)$\\
 Simple roots for $\mathfrak{t_{0}}$:\\
 \hspace*{2.5cm} compact simple roots only

\item $\mathfrak{su}_{1}^{(1)}(p, q), p+q=2n$\\
Vogan diagram:\\ 
\hspace*{2.5cm}$A_{2n-1}$, trivial automorphism,\\
\hspace*{2.5cm} two imaginary simple roots $e_{2n}-e_{1}, e_{p}-e_{p+1}$ \\
 $\mathfrak{t_{0}}=\mathfrak{su}(p)\oplus\mathfrak{su}(q)$\\
 Simple roots for $\mathfrak{t_{0}}$:\\
 \hspace*{2.5cm} compact simple roots only

\item $\mathfrak{so}_{-1}^{(1)}(2, 2n-1)$\\
Vogan diagram:\\ 
\hspace*{2.5cm}$B_{n}$, trivial automorphism,\\
\hspace*{2.5cm} one imaginary simple root $e_{1}-e_{2}$ \\
 $\mathfrak{t_{0}}=\mathfrak{so}(2n+3)$\\
 Simple roots for $\mathfrak{t_{0}}$:\\
 \hspace*{2.5cm} compact simple roots only

\item $\mathfrak{so}^{(1)}(2p, 2q+1),\;p+q=n$\\
Vogan diagram:\\ 
\hspace*{2.5cm}$B_{n}$, trivial automorphism,\\
\hspace*{2.5cm} one imaginary simple root $e_{p}-e_{p+1}$ \\
 $\mathfrak{t_{0}}= \begin{cases} \mathfrak{so}(4)\oplus\mathfrak{so}(2n-3), & \mbox{if } p=2 \\ \mathfrak{su}^{(1)}(4)\oplus\mathfrak{so}(2n-5), & \mbox{if } p=3\\
\mathfrak{so}^{(1}(2n)& \mbox{if } q=0\\
 \mathfrak{so}^{(1}(2p)\oplus\mathfrak{so}(2q), & \mbox{if } \mbox{else }
 \end{cases}$\\
 Simple roots for $\mathfrak{t_{0}}$:\\
 \hspace*{2.5cm} compact simple roots and \\
\hspace*{2.5cm} $\begin{Bmatrix}
 e_{p-1}+e_{p}& \mbox{when}\; p\geq 3\\
 \mbox{no}\;\; \mbox{other}\;\; &\mbox{when}\; p=2
 \end{Bmatrix}$

\item $\mathfrak{so}^{(1)}(1, 2n)$\\
Vogan diagram:\\ 
\hspace*{2.5cm}$B_{n}$, non-trivial automorphism,\\
\hspace*{2.5cm} no imaginary simple roots are painted \\
 $\mathfrak{t_{0}}=\mathfrak{so}^{(2)}(2n)$\\
 Simple roots for $\mathfrak{t_{0}}$:\\
 \hspace*{2.5cm} compact simple roots only and $-e_{2}$

\item $\mathfrak{so}_{1}^{(1)}(2, 2n-1)$\\
Vogan diagram:\\ 
\hspace*{2.5cm}$B_{n}$, trivial automorphism,\\
\hspace*{2.5cm} two imaginary simple roots are painted $e_{1}-e_{2},\;-e_{1}-e_{2}$ \\
 $\mathfrak{t_{0}}=\mathfrak{so}(2n+1)$\\
 Simple roots for $\mathfrak{t_{0}}$:\\
 \hspace*{2.5cm} compact simple roots only 

\item $\mathfrak{so}^{(1)}(2p+1, 2q),\;p+q=n$\\
Vogan diagram:\\ 
\hspace*{2.5cm}$B_{n}$, non-trivial automorphism,\\
\hspace*{2.5cm} one imaginary simple root $e_{p}-e_{p+1}$ \\
 $\mathfrak{t_{0}}= \begin{cases} \mathfrak{su}(3)\oplus\mathfrak{so}(2n-3), & \mbox{if } p=2 \\ \mathfrak{so}^{(2)}(2n), & \mbox{if } q=2\\
 \mathfrak{so}^{(2)}(2p)\oplus\mathfrak{so}(2q+1), & \mbox{if } \mbox{else }
 \end{cases}$\\
 Simple roots for $\mathfrak{t_{0}}$:\\
 \hspace*{2.5cm} compact simple roots and $-e_{2}, e_{p}$

\item $\mathfrak{sp}^{(1)}(p, q),\;p+q=2n \;\mbox{or}\;p+q=2n-1$\\
Vogan diagram:\\ 
\hspace*{2.5cm}$C_{n}$, trivial automorphism,\\
\hspace*{2.5cm} $p^{\mbox{th}}$ simple root painted $e_{p}-e_{p+1},$ \\
 $\mathfrak{t_{0}}=\begin{cases}
 \mathfrak{su}^{(1)}(2)\oplus \mathfrak{sp}(q) & \mbox{if } p=1, \forall q \\
 \mathfrak{sp}^{(1)}(p)\oplus\mathfrak{sp}(q) & \mbox{if } p>1, \forall q\\
 \end{cases}$\\
 Simple roots for $\mathfrak{t_{0}}$: compact simple roots and $2e_{p}$

\item $\mathfrak{sp}_{-1}^{(1)}(n, \mathbb{R})$\\
Vogan diagram:\\ 
\hspace*{2.5cm}$C_{n}$, trivial automorphism,\\
\hspace*{2.5cm} $n^{\mbox{th}}$ simple root painted $2e_{n},$ \\
 $\mathfrak{t_{0}}=\mathfrak{sp}(n)$\\
 Simple roots for $\mathfrak{t_{0}}$: compact simple roots only

\item $\mathfrak{sp}_{1}^{(1)}(n, \mathbb{R})$\\
Vogan diagram:\\ 
\hspace*{2.5cm}$C_{n}$, trivial automorphism,\\
\hspace*{2.5cm} $n^{\mbox{th}}$ and affine  simple roots are painted $2e_{n}, -2e_{1}$ \\
 $\mathfrak{t_{0}}=\mathfrak{su}(n)$\\
 Simple roots for $\mathfrak{t_{0}}$: compact simple roots only

\item $\mathfrak{sp}^{(1)}(2n-1, \mathbb{R})$\\
Vogan diagram:\\ 
\hspace*{2.5cm}$C_{n}$, non-trivial automorphism,\\
\hspace*{2.5cm} no imaginary simple root \\
 $\mathfrak{t_{0}}=\mathfrak{su}^{(2)}(2n)$\\
 Simple roots for $\mathfrak{t_{0}}$: \\
\hspace*{2.5cm}$(e_{2n-1}-e_{1})$ and \\
 \hspace*{2.5cm}all $\frac{1}{2}(e_{i}-e_{i+1}+e_{2n-1-i}-e_{2n-i})$ for $1\leq i \leq (n-1)$

\item $\mathfrak{sp}^{(1)}(n, \mathbb{H})$\\
Vogan diagram:\\ 
\hspace*{2.5cm}$C_{n}$, non-trivial automorphism,\\
\hspace*{2.5cm} no imaginary simple root \\
 $\mathfrak{t_{0}}=\mathfrak{su}^{(2)}(2n)$\\
 Simple roots for $\mathfrak{t_{0}}$: \\
\hspace*{2.5cm}$(e_{n}-e_{n+1}),(e_{2n}-e_{1})$ and \\
 \hspace*{2.5cm}all $\frac{1}{2}(e_{i}-e_{i+1}+e_{2n-i}-e_{2n+1-i})$ for $1\leq i \leq (n-1)$

\item $\mathfrak{sp}^{(1)}(2n, \mathbb{R})$\\
Vogan diagram:\\ 
\hspace*{2.5cm}$C_{n}$, non-trivial automorphism,\\
\hspace*{2.5cm} unique imaginary simple root $e_{n}-e_{n+1}$ painted\\
 $\mathfrak{t_{0}}=\mathfrak{su}^{(2)}(2n)$\\
Simple roots for $\mathfrak{t_{0}}$:\\
\hspace*{2.5cm}$\frac{1}{2}(e_{n-1}+e_{n}-e_{n+1}-e_{n+2})$ and \\
 \hspace*{2.5cm}all $\frac{1}{2}(e_{i}-e_{i+1}+e_{2n-i}-e_{2n+1-i})$ for $1\leq i \leq (n-1)$

\item $\mathfrak{so}^{(1)}(2p,2q),\; p+q=n$\\
Vogan diagram:\\ 
\hspace*{2.5cm}$D_{p+q}$, trivial automorphism,\\
\hspace*{2.5cm} $p^{\mbox{th}}$ simple root painted $e_{p}-e_{p+1},$\\
 $\mathfrak{t_{0}}=\begin{cases} 
 \mathfrak{su}^{(1)}(4)\oplus\mathfrak{so}(2n-6), & \mbox{if } p=3\\ \mathfrak{so}^{(1)}(2p)\oplus\mathfrak{so}(2q), & \mbox{if } p>3
 \end{cases}$\\
Simple roots for $\mathfrak{t_{0}}$:\\
\hspace*{2.5cm} compact simple roots and \\
\hspace*{2.5cm} $\begin{Bmatrix}
 e_{p-1}+e_{p}& \mbox{when}\; p\geq 2\\
 \mbox{no}\;\; \mbox{other}\;\; &\mbox{when}\; p=1
 \end{Bmatrix}$

\item $\mathfrak{so^{*}}^{(1)}(2n)$\\
Vogan diagram:\\ 
\hspace*{2.5cm}$D_{n}$, trivial automorphism,\\
\hspace*{2.5cm} two imaginary simple roots $(e_{1}-e_{2}),(e_{n-1}-e_{n})$ painted\\
 $\mathfrak{t_{0}}=\mathfrak{su}(n)$\\
Simple roots for $\mathfrak{t_{0}}$: compact simple roots only 

\item $\mathfrak{so}_{-1}^{(1)}(2,2n-2)$\\
Vogan diagram:\\ 
\hspace*{2.5cm}$D_{n}$, trivial automorphism,\\
\hspace*{2.5cm} two imaginary simple roots $(e_{1}-e_{2}),-(e_{1}-e_{2})$ painted\\
 $\mathfrak{t_{0}}=\mathfrak{su}(2n-2)$\\
Simple roots for $\mathfrak{t_{0}}$: compact simple roots only

\item $\mathfrak{so}_{\sigma_{v}}^{(1)}(1,2n-1)$\\
Vogan diagram:\\ 
\hspace*{2.5cm}$D_{n}$, nont-rivial automorphism,\\
\hspace*{2.5cm} no imaginary simple root painted\\
 $\mathfrak{t_{0}}=\mathfrak{so}^{(2)}(2n-2)$\\
Simple roots for $\mathfrak{t_{0}}$:
$-e_{2},\; e_{n-1}$ and all $(e_{i}-e_{i+1})$ for $2\leq i \leq (n-2)$  

\item $\mathfrak{so}_{\sigma_{v}}^{(1)}(2p+1,2q+1),\; p+q=n-1$\\
Vogan diagram:\\ 
\hspace*{2.5cm}$D_{n}$, non-trivial automorphism,\\
\hspace*{2.5cm} $p^{\mbox{th}}$ simple root painted $e_{p}-e_{p+1},$\\
 $\mathfrak{t_{0}}=\begin{cases}\mathfrak{su}(3)\oplus\mathfrak{so}(2n-5) , & \mbox{if } p=2\\
 \mathfrak{so}^{(2)}(2p)\oplus\mathfrak{so}^{(1)}(2q+1), & \mbox{if } p\geq 3\\ 
 \end{cases}$\\
Simple roots for $\mathfrak{t_{0}}$:\\
\hspace*{2.5cm} $-e_{2},\; e_{n-1},\; e_{p}$ and \\
\hspace*{2.5cm} $e_{i}-e_{i+1}$ for $2\leq i \leq p-1$ and $p+1\leq i \leq n-2$

\item $\mathfrak{so}_{\gamma}^{(1)}(1,2n-1)$\\
Vogan diagram:\\ 
\hspace*{2.5cm}$D_{n}$, non-trivial automorphism,\\
\hspace*{2.5cm} no imaginary simple root painted\\
 $\mathfrak{t_{0}}=\mathfrak{so}^{(1)}(2n-1)$\\
Simple roots for $\mathfrak{t_{0}}$:\\
\hspace*{2.5cm}$(e_{1}-e_{2}), -(e_{1}+e_{2}),\; e_{n-1}$ and \\
\hspace*{2.5cm}all $(e_{i}-e_{i+1})$ for $2\leq i \leq (n-2)$

\item $\mathfrak{so}_{\gamma}^{(1)}(2p+1,2q+1),\; p+q=n-1$\\
Vogan diagram:\\ 
\hspace*{2.5cm}$D_{n}$, non-trivial automorphism,\\
\hspace*{2.5cm} $p^{\mbox{th}}$ simple root painted $e_{p}-e_{p+1},$\\
 $\mathfrak{t_{0}}=\begin{cases} \mathfrak{so}^{(2)}(2n), & \mbox{if } p=1\\
 \mathfrak{sp}^{(1)}(2)\oplus \mathfrak{so}(2q+1), & \mbox{if } p=2\\
 \mathfrak{so}^{(1)}(2p+1)\oplus\mathfrak{so}(2q+1), & \mbox{if } p\geq3\\ 
 \end{cases}$\\
Simple roots for $\mathfrak{t_{0}}$:\\
\hspace*{2.5cm} For $p=1$\;\; $e_{p},-(e_{1}+e_{2}), e_{n-1}, $\\
\hspace*{2.5cm}all $(e_{i}-e_{i+1})$ for $p+1\leq i \leq (n-2)$\\
\hspace*{2.5cm}For $p\neq 1$\;\;$(e_{1}-e_{2}), -(e_{1}+e_{2}),\; e_{n-1},\; e_{p}$ and \\
\hspace*{2.5cm}all $(e_{i}-e_{i+1})$ for $2\leq i \leq (p-1)$ and $p+1\leq i \leq (n-2)$

\item $\mathfrak{so}_{\sigma_{s}}^{(1)}(1,2n-1), n\;\;\mbox{even}$\\
Vogan diagram:\\ 
\hspace*{2.5cm}$D_{n}$, non-trivial automorphism,\\
\hspace*{2.5cm} no imaginary simple root painted\\
 $\mathfrak{t_{0}}=\mathfrak{su}^{(2)}(n)$\\
Simple roots for $\mathfrak{t_{0}}$:\\
\hspace*{2.5cm} $\frac{1}{2}(e_{1}-e_{2}+e_{n-1}-e_{n}),\;\frac{1}{2}(-e_{1}-e_{2}+e_{n-1}+e_{n}),\;(e_{\frac{n}{2}}-e_{\frac{n}{2}+1})$ and \\
 \hspace*{2.5cm}all $\frac{1}{2}(e_{i}-e_{i+1}+e_{n-i}-e_{n+1-i})$ for $2\leq i \leq (n-2)$

\item $\mathfrak{so}_{\sigma_{s}}^{(1)}(1,2n-1), n\;\; \mbox{even}$\\
Vogan diagram:\\ 
\hspace*{2.5cm}$D_{n}$, non-trivial automorphism,\\
\hspace*{2.5cm} unique imaginary simple root $(e_{\frac{n}{2}}-e_{\frac{n}{2}+1})$ painted\\
 $\mathfrak{t_{0}}=\mathfrak{so}^{(1)}(n)$\\
Simple roots for $\mathfrak{t_{0}}$:\\
\hspace*{2.5cm} $\frac{1}{2}(e_{1}-e_{2}+e_{n-1}-e_{n}),\;\frac{1}{2}(-e_{1}-e_{2}+e_{n-1}+e_{n}),\\\hspace*{2.5cm}\frac{1}{2}(e_{\frac{n}{2}-1}+e_{\frac{n}{2}}-e_{\frac{n}{2}+1}-e_{\frac{n}{2}+2})$ and \\
 \hspace*{2.5cm}all $\frac{1}{2}(e_{i}-e_{i+1}+e_{n-i}-e_{n+1-i})$ for $2\leq i \leq (n-2)$

\item $\mathfrak{so}_{\sigma_{s}}^{(1)}(1,2n-1), n \mbox{odd}$\\
Vogan diagram:\\ 
\hspace*{2.5cm}$D_{n}$, non-trivial automorphism,\\
\hspace*{2.5cm} no imaginary simple root painted\\
 $\mathfrak{t_{0}}=\mathfrak{so}^{(1)}(n)$\\
Simple roots for $\mathfrak{t_{0}}$:\\
\hspace*{2.5cm} $\frac{1}{2}(e_{1}-e_{2}+e_{n-1}-e_{n}),\;\frac{1}{2}(-e_{1}-e_{2}+e_{n-1}+e_{n})$, and \\
 \hspace*{2.5cm}all $\frac{1}{2}(e_{i}-e_{i+1}+e_{n-i}-e_{n+1-i})$ for $2\leq i \leq (n-2)$
\end{itemize}
\subsection{Notations used}
\begin{itemize}
 \item $gl(n, \mathbb{C}),(gl(n, \mathbb{R}))$: \{all $n\times n$ complex (real matrices)\}
 \item $sl(n, \mathbb{C}),(sl(n, \mathbb{R}))$: \{all $n\times n$ complex (real matrices) of trace zero\}
 \item $sl_{1}^{(1)}(n,\mathbb{R})$: $sl(n, \mathbb{R}) \otimes \mathbb{C}[t, t^{-1}]\oplus \mathbb{R}ic\oplus \mathbb{R}id$ with $u=1$
  \item $sl_{-1}^{(1)}(n,\mathbb{R})$: $sl(n, \mathbb{R}) \otimes \mathbb{C}[t, t^{-1}]\oplus \mathbb{R}ic\oplus \mathbb{R}id$ with $u=-1$
 \item $so(n):\{X\in gl(n, \mathbb{R})\mid X+X^{*}=0 \}$
 \item $so_{1}^{(1)}(n)$: $so(n)\otimes\mathbb{C}[t,t^{-1}]\oplus \mathbb{R}ic\oplus \mathbb{R}id$ with $u=1$
 \item $so_{-1}^{(1)}(n)$: $so(n)\otimes\mathbb{C}[t,t^{-1}]\oplus \mathbb{R}ic\oplus \mathbb{R}id$ with $u=-1$
 \item $sp(n):\{X\in gl(n, \mathbb{H})\mid X+X^{*}=0 \}$
 \item $su(n):\{X\in gl(n, \mathbb{C})\mid X+X^{*}=0$ and $Tr X=0 \}$
 \item $su^{(1)}(n):\{X\in su(n)\otimes\mathbb{C}[t,t^{-1}]\oplus \mathbb{R}ic\oplus \mathbb{R}id\mid X+X^{*}=0$ and $Tr X=0 \}$
 \item $su^{(2)}(n):{\displaystyle\sum_{p=0}^{1}}{\displaystyle \sum_{j\; mod\; 2=p}}t^{j}\otimes su(n)_{p}^{(2)}\oplus \mathbb{R}ic\oplus \mathbb{R}id$ where $su(n)_{0}^{(2)}$ and  $su(n)_{1}^{(2)}$ are the eigenspaces corresponding to eigenvalues $1$ and $e^{\pi i}$ respectively. That is $a \in su(n)_{p}^{(2)}$ if $a \in su(n)$ and $\Psi_{\tau} (a)=e^{\pi ip}$ where$\Psi_{\tau}$ is the is the outer automorphism of $su(n).$ 
   \item $u(p,q): \Bigg\{\begin{pmatrix}
               Z_{1} & Z_{2}\\ \overline{Z}_{2}^{t} & Z_{3}
                \end{pmatrix} \mid  \begin{array}{c} Z_{1}, Z_{3}$ skew Hermitian of order $p$ and $q$ $ \\
$ respectively, $Z_{2}$  is arbitrary$  \end{array} \Bigg\}$
 \item $su(p,q): \Bigg\{\begin{pmatrix}
               Z_{1} & Z_{2}\\ 
               \overline{Z}_{2}^{t} & Z_{3}
                \end{pmatrix} \mid  \begin{array}{c} Z_{1}$ skew Hermitian of order $p$ $ \cr
  Z_{3}$ skew Hermitian of order $q$ $ \cr
  Tr Z_{1}+Tr Z_{3}=0$, $Z_{2}$  is arbitrary$  \end{array} \Bigg\}$
\item $su_{1}^{(1)}(p,q):\Bigg\{\begin{pmatrix}
               X_{1} & X_{2}\\ \overline{X}_{2}^{t} & X_{3}
                \end{pmatrix} \mid  \begin{array}{c} X_{i}=Z_{i}\otimes\mathbb{C}[t,t^{-1}]$  $\cr
X_{1}$ skew Hermitian of order $p$ $ \\
  X_{3}$ skew Hermitian of order $q$ $ \\
  Tr X_{1}+Tr X_{3}=0$ $X_{2}$  is arbitrary$  \end{array} \Bigg\}$
  \item $su_{-1}^{(1)}(p,q):\Bigg\{\begin{pmatrix}
               X_{1} & X_{2}\\-\overline{X}_{2}^{t} & X_{3}
                \end{pmatrix} \mid  \begin{array}{c} X_{i}=Z_{i}\otimes\mathbb{C}[t,t^{-1}]$  $\cr
X_{1}$ skew Hermitian of order $p$ $ \\
  X_{3}$ skew Hermitian of order $q$ $ \\
  Tr X_{1}+Tr X_{3}=0$, $X_{2}$  is arbitrary$  \end{array} \Bigg\}$
  \item $su^{\ast}(2n): \Bigg\{\begin{pmatrix}
               Z_{1} & Z_{2}\\ -\overline{Z}_{2} & \overline{Z_{1}}
                \end{pmatrix} \mid  \begin{array}{c} Z_{1}, Z_{2}$ $n\times n$ complex matrix $ \cr
Tr Z_{1}+ Tr\overline{Z_{1}}=0  \end{array} \Bigg\}$
\item $sp(n, \mathbb{C}): \Bigg\{\begin{pmatrix}
               Z_{1} & Z_{2}\\ \overline{Z}_{3} & -Z_{1}^{t}
                \end{pmatrix} \mid  \begin{array}{c} Z_{i}$ $n\times n$ complex matrix $ \cr
Z_{2}, Z_{2}$  are symmetric$  \end{array} \Bigg\}$
\end{itemize}


\begin{thebibliography}{}
\bibitem{Batra2000} P. Batra, Invariants of real forms of affine Kac-Moody Lie algebras, {\it J. Algebra} {\textbf 223} (2000), 208--236.
\bibitem{Batra2002}P. Batra Vogan diagrams of real forms of affine Kac-Moody Lie algebras, {\it J. Algebra} {\textbf 251} (2002), 80--97.
\bibitem{Birkhoff1937} G.Birkhoff, Representation of Lie Algebras and Lie Groups by Matrices,  {\it Annals of Mathematics} \textbf{38}(1937), 526-532.
\bibitem{Caselle} M. Caselle, and  U. Magnea, Random matrix theory and symmetric spaces, {\it arxiv: cond-mat/0304363} \textbf{v 2}.
\bibitem{Caselle2006} M.Caselle, and U. Magnea, Symmetric space description of carbon nano tube., {\it arXiv: Cond-mat/0506733}\textbf{v 2}.
\bibitem{Chuah2004} M.K. Chuah, and  C.C. Hu., Equivalence classes of Vogan diagram, {\it J. Algebra}\textbf{279}, (2004) 22-37.
\bibitem{Chuah2006} M.K. Chuah, and  C.C. Hu., Extended Voagn diagrams, {\it J. Algebra}\textbf{301}, (2006) 112--147.
\bibitem{Cornwell1992} J. F. Cornwell, General theory of the matrix formulation of the automorphism of affine Kac-Moody algebras. {\it J. Phys. A.: Math. Gen} {\textbf 25} (1992),2311-2333.
\bibitem{Dyson1970} F. Dyson, {\it Comm. Math. Phys} {\textbf 19}(1970), 235. 
\bibitem{Freyn2009} W.Freyn, Kac-Moody symmetric spaces and universal twin buildings, PhD thesis, Universitat Augsburg, 2009.
\bibitem{Fuchs1996} J.Fuchs, B. schellekens, and C. Schweigert {\it Comm. Math. Phys} {\textbf 180}(1996), 39-97. 
\bibitem{Gilmore}R. Gilmore, Lie groups, Lie algebras, and some of their applications, Dover Publications, INC.
\bibitem{Heintze2006} E. Heintze,  Toward symmetric spaces of affine Kac-Moody type, {\it Int. J. Geo. Methods in Modern Physics}, {\textbf 3} No.s 45 and 6(2006), 881-898.
\bibitem{Heintze2012} E. Heintze, and G.Grob, Finite order automorphisms and real forms of affine Kac-Moody algebras in the smooth and algebraic category, {\it Memories of the American mathematical society}, {\textbf 219}(2012), 1030.
\bibitem{Kac1968}  V.G. Kac, Simple irreducile graded Lie algebras of finite growth(russian), {\it Math. USSR-Izvestiya} {\textbf 2}(1968), 1271-1311.
\bibitem{Kac1990} V. G. Kac, Infinite dimensional Lie algebras, 3rd edition, {\it Cambridge University Press}(1990).
\bibitem{Kobayashi1986} Z. Kobayashi, Automorphisms of finite order of the affine Lie algebra $A_{l}^{(1)}$, {\it Tsukuba J. Math.}, Vol. 10, No. 2(1986), 269-283.
\bibitem{Knapp2002} A. W. Knapp, Lie Groups Beyond an Introduction, Second Edition, {\it Birkh$\ddot{a}$user}.
\bibitem{Levstein1988} F. Levstein, A Classification Of Involutive automorphism of an Affine Kac-Moody Lie Algebra, {\it J. of Algebra} {\textbf 114}(1988), 489-518.
\bibitem{Loos 1969} O. Loos, Symmetric spaces II: Compact spaces and classification, {\it Benjamin} (1969), MR0239006(39:365b).
\bibitem{Moody1967} R.V.  Moody, Lie algebra associated with generalized Cartan matrices, {\it Bull. Amer.Math.Soc.} {\textbf 73}(1967),217-221.
\bibitem{Olshanetsky1983} M.A. Olshanetsky, and  A.M. Perelomov, {\it Phys. Rep.} {\textbf 94}(1983), 313.
\bibitem{Popescu2005} B. Popescu, Infinite dimensional symmetric spaces, Thesis, University of Augsburg, (2005).
\bibitem{Rousseau2003}G. Rousseau, and  H.B. Messaoud, Classifications desformes reelles presque compactes des algebras de Kac-Moody affines, {\it J. of Algebra}, (2003), 443-513.
\bibitem{Helgason2001}  H. Sigurdur, Differential Geometry, Lie Groups, and Symmetric Spaces, {\it AMS} Volume- {\textbf 34} (2001).
\bibitem{Terng1995}  Chuu-Lian Terng, Polar actions on Hilbert space, {\it J. of Geon} {\textbf 5(1)} (1995), 129-150.
\bibitem{Tripathy2006}  L. K. Tripathy, and  K.C. Pati, Satake diagrams of affine Kac-Moody algebras, {\it J. Phys. A: Math Gen.} {\textbf 39}(2006), 1385.
\bibitem{Jacobson1979} N. Jacobson, Lie Algebras, {\it Dover}, New York(1979).
\end{thebibliography}
\end{document}